\def\erfc{\hbox{\rm erfc}}

\documentclass[12pt,a4paper,onecolumn]{article}
\begin{document}

\title{Diffraction of Electromagnetic Waves}

\author{Ulrich Brosa\footnote{brosa-gmbh@t-online.de}\\
        Brosa GmbH, Am Br\"ucker Tor 4, D-35287 Am\"oneburg, Germany\\
        and Philipps-Universit\"at, Renthof 6, D-35032 Marburg}
\maketitle
\begin{abstract}
The general method to obtain solutions of the Maxwellian equations
from scalar representatives is developed and applied to the diffraction
of electromagnetic waves. Kirchhoff's integral is modified to provide
explicit expressions for these representatives. The respective integrals
are then evaluated using the method of stationary phase in two dimensions.
Hitherto unknown formulae for the polarization appear as well as
for imaging by diffraction. Ready-to-use formulae describing Fresnel
diffraction behind a round stop are presented.



\bigskip\noindent
(The final version appeared in Z.~Naturforsch.~{\bf 65a},~1-24~(2010).)
\end{abstract}
\section{Sad state of theory}
\label{sec:sad}
The first impetus to this work came when your author read an altogether
reasonable preprint on the Casimir effect.  Its author, though obviously
being a skilled physicist, wrestled heftily to establish the normal modes
of the electromagnetic field between two metallic plates. The
modified preprint is now published \cite{Sch08}.

The second impetus came when your author lectured on theoretical optics.
Diffraction of light belongs to the canon. He explained
to the students what he had found in the treatise of Nobelist
Max Born \cite{Bor65} and realized only after a few weeks how bad it is.
Born's book and its enhancement \cite{Bor97} are subtitled
{\lq\lq}Electromagnetic Theory of Propagation, Interference and Diffraction
of Light{\rq\rq}. A truthful subtitle should rather be
{\lq\lq}A Wrong Theory of Sound{\rq\rq}, at least with respect to diffraction.

Nobody, neither physicist nor mathematician, seems to be able to
systematically solve partial differential equations for vector fields.
But the Maxwell equations, foundation of optics and quantum electrodynamics,
are just equations of that kin. As a consequence, the theory of diffraction
is still a graveyard of intellectual atrocities.

{\lq\lq}Graveyard{\rq\rq} in German means {\lq\lq}Kirchhof{\rq\rq}.
It is Gustav Kirchhoff who is held responsible for the theory
stolid till now \cite{Kir82},\cite{Kir83},\cite{Kir91}.
Even in most recent textbooks Kirchhoff's mistakes are recited
as {\it the\/} theory of diffraction
\cite{Hec02},\cite{Mes04},\cite{Sha06},\cite{Kon08},
see \cite{Hoe61} for a historic review.

Kirchhoff was aware of Maxwell's work, but did not esteem it too much.
He preferred a product made in Germany, namely the Helmholtz equation
\begin{equation}
\nabla_p^2\ \psi({\bf r}_p)+k^2\,\psi({\bf r}_p)=0             \label{hel1}
\end{equation}
wherein ${\bf r}_p$ is to denote the place of the probe,
$\nabla_p$ the nabla operator acting on ${\bf r}_p$ and
$k$ the wave number. $\psi$ is the world-reputed symbol of parapsychology.
Nobody knows what it means. Kirchhoff baptized it
{\lq\lq}Verr\"uckung eines Aethertheilchens{\rq\rq} translatable
as {\lq\lq}insanity of an etheral piece{\rq\rq} \cite[p.9]{Kir91}.

Today Maxwell's equations are not despised that much.
It is generally accepted that light is an electromagnetic wave.
Nevertheless scientists, even highly qualified ones, stick to
Kirchhoff's doctrines. They observe sagaciously
that in Cartesian coordinates one may extract from Maxwell's equations
one Helmholtz equation for each and every Cartesian component of the
electric field ${\bf E}$ and the magnetic one ${\bf B}$ \cite[Chap.3]{Hec02}.
The value of this observation is zero. The components of the
electromagnetic vectors are coupled through the Maxwellian equations
for curls and divergences. Solving the six Helmholtz equations for the
components of ${\bf E}$ and ${\bf B}$ independently generates grossly
wrong results. Even worse: In problems where Cartesian coordinates
do not suit, as diffraction by spheres, there is,
for the components of ${\bf E}$ and ${\bf B}$, no Helmholtz
equation at all \cite{Mie08}.

The next row of atrocities was digged when Kirchhoff, following Hermann
Helmholtz, considered an integral which, seemingly, solves the Helmholtz
equation \cite[p.82]{Kir91}
\begin{equation}
\psi({\bf r}_p)={1\over 4\pi}\int\!\!\int_F\bigg(
{\exp(ik|{\bf r}_p-{\bf r}|)\over|{\bf r}_p-{\bf r}|}\ \partial_n\psi({\bf r})
-\psi({\bf r})\ \partial_n{\exp(ik|{\bf r}_p-{\bf r}|)\over|{\bf r}_p-{\bf r}|}
\bigg)\ \hbox{d}f                                                \label{kir1}
\end{equation}
wherein the surface $F$ is to divide the entire three-dimensional
space in an inner and an outer part; the inner part is where the probe
${\bf r}_p$ resides. ${\bf r}$ points to a point on $F$; it incorporates
the variables of integration as components. $\hbox{d}f$ is the respective
element of the surface. $\partial_n$ symbolizes a differentiation in the
direction of the outer normal on $F$. The integral is interpreted
in a way that primary waves $\psi({\bf r})$ travel
through outer space until they strike $F$; there they excite secondary
spherical waves $\exp(ik|{\bf r}_p-{\bf r}|)/|{\bf r}_p-{\bf r}|$
interfering to produce the wanted $\psi({\bf r}_p)$.

Arnold Sommerfeld has criticized that simultaneous fixing of
boundary values for the function and its derivative as required
in Kirchhoff's integral (\ref{kir1}) causes contradictions
\cite{Som96},\cite{Som64},\cite{Som64e}. But scientists answered
and still answer, they would take for $\psi({\bf r})$ a plane wave
and differentiate it consistently, without contradiction.
These scientists do not understand that diffraction is possible
only if part of the primary wave is screened. How should one treat
these parts of $F$? Kirchhoff and his followers demand simultaneously
$\psi({\bf r})=0$ and $\partial_n\psi({\bf r})=0$ on the screened part
of $F$ and call it a {\it black screen\/} \cite[p.40]{Kir91}.
Yet the Helmholtz equation is elliptic of second degree.
It has no real characteristics. Hence it follows from the Cauchy Kovalevskaya
Theorem that the only solution compatible with a {\it black screen\/} is a
global zero. Kirchhoff's integral produces no better
than order-of-magnitude results.

The third row of atrocites was graven when Kirchhoff
wanted to evaluate the integral (\ref{kir1}), but could not do it exactly.
Especially the function in the exponents seemed
to be invincible. Kirchhoff did then what all physicists do
when they lack ideas. He substituted for the invincible function
its crippled Taylor expansion. The evaluation with linear terms in
${\bf r}$ only was called Fraunhofer diffraction, whereas truncation
after quadratic terms in ${\bf r}$ was said to yield Fresnel
diffraction \cite[p.86]{Kir91}.

One should assume that this kind of Fresnel diffraction
includes Fraunhofer diffraction as the special case in which quadratic
terms are negligible. But this is not true. The formulae derived in this
way describe fundamentally different pattern.

All these problems will be solved in the present article.

The general method to decouple Maxwell's equations
will be developed in section \ref{sec:representatives}.
The electromagnetic field is represented by three scalar functions
$a$, $b$ and $c$ which obey separated differential equations.
One may solve the separated equations for these mathematical auxiliaries
one by one and afterwards calculate the physical fields ${\bf E}$ and ${\bf B}$
from the auxiliaries by straightforward differentiation.
$c$ is the familiar scalar potential, $a$ and $b$ are common
factors in the components of the simple and the double vector potential,
respectively. This will be demonstrated for all
electrodynamics in homogeneous and isotropic materials, in particular
for those with finite conductivity. The General Representation
Theorem in section \ref{sec:representatives} constitutes the first main
result in this article.

In electrodynamics, diffraction is just a simple special case
for which the representatives $a$ and $b$ suffice.

In all problems with partial differential equations, also in the theory
of diffraction, suitable boundary values must be fixed.
Instead of the inconsistent {\it black screen\/} we will put up
a screen of perfect conductivity. This entails boundary conditions
for the electrical field ${\bf E}$. When these are converted for
$a$ and $b$, it turns out that the one representative obeys a
homogeneous boundary condition of the first kind (aka Dirichlet),
whereas the other is subject to a homogenous boundary condition
of the second kind (aka Neumann), see section \ref{sec:boundary}.

To complete the mathematical definition of diffraction,
initial values must be given. In section \ref{sec:initial} we will account
for them using a Laplace transform. Then the transformed representatives
$a$ and $b$ fulfill separated Helmholtz equations. This explains why
Kirchhoff's formulas are not entirely wrong.

For the representation theorem the way the representatives
are found does not matter. Expansions in terms of partial waves are,
for example, feasible, but the aim of this article is
to mend Kirchhoff's advances. Since the representatives $a$ and $b$
fulfill Helmholtz equations, one may derive for them integrals
similar to (\ref{kir1}). As Sommerfeld's criticism must be attended,
the spherical wave will be replaced with genuine Green functions
for boundary problems of the first and the second kind
in section \ref{sec:kirchhoff}.

Hence it might seem that, for the true theory of electromagnetic
diffraction, expense is double as two integrals must be computed.
Yet if for the primary wave
a spherical wave is appointed that proceeds from a source at
${\bf r}_q$, both $a$ and $b$ are determined by the same function
of ${\bf r}_p$ and ${\bf r}_q$, the only difference being that
probe and source are interchanged. The equations (\ref{kir13})
through (\ref{kir16}) display the second main result in this
article.

The invincible function in the integral, mentioned above,
can be rewritten to have an intuitive meaning:
It is the difference of lengths when the points
${\bf r}_q$ and ${\bf r}_p$  are connected either directly or
via a point ${\bf r}$ on the screen. These very terms occur
in the familiar triangle inequation. It follows from
the properties of the triangle function that simple geometry determines
the basics of diffraction. For example, the border of shadow
can be derived from the zero of the triangle function. The Criterion
of Light will be established in section \ref{sec:triangle}.

Augustin Fresnel introduced, for the description of diffraction,
certain integrals. For a realistic theory of diffraction these integrals
are too clumpsy. It is more convienient to use instead the
error function known to all statisticians. Although one needs, for
diffraction, the error function of a complex argument,
it behaves in a similar way as that of a real argument.
Aside for greater simplicity, we get from the error function
more results, namely diffraction in dissipative materials,
which is of paramount importance for practical purposes.
The salient properties of the complex error function will be
described in section \ref{sec:error}.

There are several methods to evaluate the integral (\ref{kir14}),
e.g.\ for great distances of the probe from the screen.
In this article we will apply the method of stationary phase
to find the integral in the limit of short waves.
It is this what Kirchhoff wanted when he expanded his functions
up to second powers. We will avoid Taylor's expansion and
introduce instead new variables that map the triangle function exactly.
How this has to be done for two- and higher-dimensional integrals with
finite limits, was hitherto unknown. It requires a novel deliberation.
The Principle of Utter Exhaust will be introduced in section \ref{sec:exhaust}.
It constitutes the third main result in this article.

Utter exhaust (\ref{exh5}) will be applied to the corrected Kirchhoff
integral (\ref{kir14}) in section \ref{sec:universal}. It yields
the fourth main result in this article, the Universal
Formula of Diffraction (\ref{uni9}): The diffracted wave equals the
primary wave times the complementary error function which accounts
through its argument for the specific shape of the edge;
the argument is found by a purely algebraic calculation.
The formula is universal in so far as it holds
for all single diffracting edges. The diffraction by screens with
several edges can be derived from it by mere superposition.
The universal formula also holds for probes arbitrarily close
to the screen. Therefore it can describe the transition
from Fresnel to Fraunhofer diffraction. It also holds for
dissipative materials; the argument of the error
function also accounts for damping. Yet the formula (\ref{uni9})
does not describe diffraction of very long waves. Moreover we will
notice, under certain conditions, its weakness for great distances
of the probe from the screen.

In this article, no graph of calculated fields will be shown. Detailed
comparison to or prediction of measurements is here out of
scope. Nevertheless some impressive applications will be outlined.
The first is diffraction by a straight edge in section \ref{sec:edge}.
Diffraction means breaking beams asunder. Nevertheless a metallic
half screen creates via diffraction an image of the source,
that is focusing of beams. Astonishing as this result might appear,
the same formula contains as a limiting case Sommerfeld's stringent solution
describing the diffraction of a plain wave by a half plane, see section
\ref{sec:imaging}. Thus the asymptotic methods developed here
have more power than Sommerfeld's stringent integration of Maxwell's
equations.

From the diffraction by a single straight edge it is just
a small step to the diffraction by a slit, namely
a superposition. Nevertheless the ensuing formula
describes both Fresnel and Fraunhofer diffraction and
the transition between these regimes, see section \ref{sec:slit}.

In section \ref{sec:circular} the universal formula is applied to
diffraction by a circular aperture. It results in simple formulae
describing Fresnel diffraction behind a round stop, a device that
is used in almost all optical instruments.

In the concluding section \ref{sec:outlook} the possibly novel
results of this work will be listed and a program what is to be
done next be given. The article ends with a surprise.

This work is a hommage to the natural mind. All calculations
were done using a pencil and sheets of paper. Aid from computers
was rejected.

\section{The representatives of electrodynamics}
\label{sec:representatives}
The purpose of the representation theorem is to replace Maxwell's
vector equations for the magnetic and electric fields ${\bf B}$
and ${\bf E}$ with separated differential equations for scalar
representatives $a$, $b$ and $c$. As soon as the latter equations
are solved, successively, one may determine the vector fields from
the scalar representatives by straightforward differentiation.
To understand and to prove the general representation theorem of
electrodynamics, three requisites are needed. First, the

\medskip
\noindent{\bf Lemma of Triple Curl.}
{\sl The vector differential equation
\begin{equation}
\nabla\times\nabla\times\nabla\times{\bf v}a({\bf r},t)=
-\nabla\times D_t{\bf v}a({\bf r},t)                          \label{tricur1}
\end{equation}
can be reduced to the scalar differential equation
\begin{equation}
\nabla^2 a({\bf r},t)=D_t a({\bf r},t)+
\cases{f({\bf v}_0{\bf r},t)  &if\ $v_1=0$ \cr
       f(|{\bf v}|,t)         &otherwise \cr}                \label{tricur2}
\end{equation}
if and only if the supporting vector field ${\bf v}$ is preformed as
\begin{equation}
{\bf v}={\bf v}_0+v_1{\bf r}                                 \label{tricur3}
\end{equation}
with arbitrary constants ${\bf v}_0$ and $v_1$. $D_t$ symbolizes an
operator possibly including differentiations with respect to time $t$,
but definitely no differentiation with respect to space ${\bf r}$}.
$f(\cdot,t)$ denotes a free function.
\footnote{In this section and the ensuing two, the
index $p$ at the position of the probe ${\bf r}_p$ will be omitted.}
\medskip

$D_t$ is, for example, $\varepsilon\mu\partial_t^2+\mu\sigma\partial_t$,
see equation (\ref{con2}) below, with constants $\varepsilon$,
$\mu$ and $\sigma$ denoting dielectric constant, magnetic permeability
and conductivity, respectively. $f(\cdot,t)$ is a so-called gauge,
a function which can be chosen according to convenience. For present purposes
$f(\cdot,t)=0$ suffices.

The lemma was proven in \cite{Bro85}. It was published
in \cite{Bro86} and is available now in a textbook \cite{Gro93}.
It is useful for uncoupling all vector equations which describe
physical phenomena in a homogeneous and isotropic space,
e.g.\ in the theory of elasticity and liquidity.

Second, it is assumed that the electromagetic field propagates
in an isotropic and at least piecewise homogeneous medium.
The constitutive relations ${\bf D}=\varepsilon{\bf E}$ and
${\bf H}={\bf B}/\mu$, which relate the force-exerting
fields ${\bf E}$ and ${\bf B}$ with the source-caused fields
${\bf D}$ and ${\bf H}$, as well as Ohm's law
\begin{equation}
{\bf j}=\sigma{\bf E}\ ,                                      \label{ohm}
\end{equation}
which relates the electric field ${\bf E}$ with the electric
current density ${\bf j}$, will be mounted in Maxwell's
equations from the very start.

Third, as we want to solve initial- and boundary-value problems
in time $t$ and three-dimensional space ${\bf r}$ for the
electric field ${\bf E}({\bf r},t)$ and the magnetic field
${\bf B}({\bf r},t)$,
it is necessary to discriminate between charges and currents
that are enforced from outside and those that come about through
the free play between inside fields. The charge density $\rho_o({\bf r})$
at the initial time $t=t_o$ is determined by reasons outside the
considered system. In conducting materials it
decays exponentially. The remainder $\rho({\bf r},t)$ is zero
at $t=t_o$ and thus determined by the density of
the current because of continuity
\begin{equation}
\partial_t(\rho({\bf r},t)+
\rho_o({\bf r})\exp\bigg({\sigma\over\varepsilon}(t_o-t)\bigg))
=-\nabla({\bf j}_e({\bf r},t)+{\bf j}({\bf r},t)+{\bf j}_o({\bf r},t)).            \label{cont}
\end{equation}
The current density, in its turn, is partly generated by external sources
${\bf j}_e({\bf r},t)$. In conducting materials, moreover, the
electric field drives inner currents ${\bf j}({\bf r},t)+{\bf j}_o({\bf r},t)$
according to Ohm's law (\ref{ohm}). The latter is the current caused
by the decay of $\rho_o({\bf r})$.

\medskip
\noindent{\bf General Representation Theorem of Electrodynamics.}
{\sl Solutions of the Maxwellian equations
\begin{equation}
\nabla\times{\bf B}({\bf r},t)=\varepsilon\mu\partial_t{\bf E}({\bf r},t)
+\mu\sigma{\bf E}({\bf r},t)+\mu{\bf j}_e({\bf r},t)            \label{max1}
\end{equation}
\begin{equation}
\nabla\times{\bf E}({\bf r},t)=-\partial_t{\bf B}({\bf r},t)    \label{max2}
\end{equation}
\begin{equation}
\nabla{\bf E}({\bf r},t)={1\over\varepsilon}(\rho({\bf r},t)
+\rho_o({\bf r})\exp\bigg({\sigma\over\varepsilon}(t_o-t)\bigg))\label{max3}
\end{equation}
\begin{equation}
\nabla{\bf B}({\bf r},t)=0                                      \label{max4}
\end{equation}
are provided by the solutions of the differential equations
for the scalar representatives $a$, $b$ and $c$
\begin{equation}
\nabla^2 a({\bf r},t)=\varepsilon\mu\partial_t^2 a({\bf r},t)
+\mu\sigma\partial_t a({\bf r},t)
-\mu\int_{t_o}^t j_e({\bf r},\tau)
\exp\bigg({\sigma\over\varepsilon}(\tau-t)\bigg)\hbox{d}\tau       \label{con1}
\end{equation}
\begin{equation}
\nabla^2 b({\bf r},t)=\varepsilon\mu\partial_t^2 b({\bf r},t)
+\mu\sigma\partial_t b({\bf r},t)                               \label{con2}
\end{equation}
\begin{equation}
\nabla^2 c({\bf r})=-{1\over\varepsilon}\rho_o({\bf r})            \label{con3}
\end{equation}
if the magnetic and electric fields {\bf B} and {\bf E} are computed
from $a$, $b$ and $c$ according to
\begin{eqnarray}
{\bf B}({\bf r},t)=
\nabla\times{\bf v}\bigg({\sigma\over\varepsilon}+\partial_t\bigg)a({\bf r},t)
-\nabla\times\nabla\times{\bf v}b({\bf r},t)\quad\qquad  \label{rep1}\\
{\bf E}({\bf r},t)={1\over\varepsilon\mu}
\nabla\times\nabla\times{\bf v}a({\bf r},t)
+\nabla\times{\bf v}\partial_t b({\bf r},t)\quad\qquad\qquad\nonumber\\
-\nabla c({\bf r})\exp\bigg({\sigma\over\varepsilon}(t_o-t)\bigg)
-{1\over\varepsilon}\int_{t_o}^t {\bf j}_e({\bf r},\tau)
\exp\bigg({\sigma\over\varepsilon}(\tau-t)\bigg)\hbox{d}\tau\ .   \label{rep2}
\end{eqnarray}
The supporting vector field ${\bf v}$ must
be parallel to the density of the enforced current
\begin{equation}
{\bf j}_e({\bf r},t)={\bf v}j_e({\bf r},t)\ .                     \label{con4}
\end{equation}
For the general three-dimensional density, this amounts to choosing
three linearly independent supporting fields according to (\ref{tricur3}),
introducing three representatives $a$ and solving three
equations (\ref{con1}).}
\medskip

The equations (\ref{con1}) through (\ref{con4}) constitute
the first major item of this article.

The proof proceeds in three steps since Maxwell's equations constitute
a linear system. We compose the general solution from particular ones.
Let us begin with the extraction of nonhomogeneities.

First step: The ansatz
\begin{equation}
{\bf E}({\bf r},t)=
-\nabla c({\bf r})\exp\bigg({\sigma\over\varepsilon}(t_o-t)\bigg)
 \qquad {\bf B}({\bf r},t)={\bf 0}                             \label{ans1}
\end{equation}
is to extract the nonhomogeneity of (\ref{max3}) with $\rho_o({\bf r})$.
The ansatz satisfies all Maxwellian equations except the third (\ref{max3}).
The third yields the Poisson equation (\ref{con3}) for the scalar
potential $c({\bf r})$.
Please find ansatz (\ref{ans1}) linearly enclosed in the general
representation formulae (\ref{rep1}) and (\ref{rep2}).

The Ohmian current (\ref{ohm}) caused by this electric field
\begin{equation}
{\bf j}_o({\bf r},t)=
-\sigma\nabla c({\bf r})\exp\bigg({\sigma\over\varepsilon}(t_o-t)\bigg)
                                                                \label{ohm2}
\end{equation}
balances in (\ref{cont}) the term with the charge density
$\rho_o({\bf r})$. We discard them both to go on
with a simplified equation of continuity and observe
that it still facilitates the elimination of $\rho({\bf r},t)$.

Hence, for the remaining nonhomogeneities of Maxwell's equations,
we do not miss anything when we differentiate the remainder of (\ref{max3})
with respect to time
\begin{equation}
\nabla\big(\partial_t{\bf E}({\bf r},t)+
{\sigma\over\varepsilon}{\bf E}({\bf r},t)
+{1\over\varepsilon}{\bf j}_e({\bf r},t)\big)=0\ .               \label{max3a}
\end{equation}
The equation of continuity (\ref{cont}) was used to get rid of
the charge density $\rho({\bf r},t)$. Ohm's law (\ref{ohm})
was applied to eliminate the current density ${\bf j}({\bf r},t)$.
Equation (\ref{max3a}), however, is guaranteed if the first Maxwellian
equation (\ref{max1}) is fulfilled. To see this, one just has to take
its divergence.

We can completely forget about the third and fourth Maxwellians
(\ref{max3}-\ref{max4}) when we represent the magnetic field ${\bf B}$
by a vector potential ${\bf A}({\bf r},t)$, i.e.\
${\bf B}({\bf r},t)=\nabla\times{\bf A}({\bf r},t)$.
This shall be done henceforth. One should, however,
keep in mind that mere introduction of a vector potential does not help
much. The differential equations for the vector potential are coupled
in a similiar way as Maxwell's equations for the electric and magnetic
fields. We must construct special vector potentials to obtain uncoupled
differential equations.

Second step: The ansatz
\begin{equation}
{\bf B}({\bf r},t)=\nabla\times{\bf v}\alpha({\bf r},t)           \label{ans2}
\end{equation}
introduces, as announced, a special vector potential,
${\bf A}({\bf r},t)={\bf v}\alpha({\bf r},t)$, viz.\
a predetermined vector field times a free scalar function.

It shall be used to extract the nonhomogeneity ${\bf j}_e({\bf r},t)$.
The first Maxwellian (\ref{max1}) can be written as
\begin{equation}
\partial_t{\bf E}({\bf r},t)+{\sigma\over\varepsilon}{\bf E}({\bf r},t)+
{1\over\varepsilon}{\bf j}_e({\bf r},t)={1\over\varepsilon\mu}
\nabla\times\nabla\times{\bf v}\alpha({\bf r},t)\ .            \label{max1a}
\end{equation}
Integrating this equation with respect to time yields
a representation of the electric field
\begin{eqnarray}
{\bf E}({\bf r},t)={1\over\varepsilon\mu}
\nabla\times\nabla\times{\bf v}\int_{t_o}^t\alpha({\bf r},\tau)
\exp\bigg({\sigma\over\varepsilon}(\tau-t)\bigg)\hbox{d}\tau\    \nonumber\\
-{1\over\varepsilon}\int_{t_o}^t j_e({\bf r},\tau)
\exp\bigg({\sigma\over\varepsilon}(\tau-t)\bigg)\hbox{d}\tau\,.  \label{ans3}
\end{eqnarray}
Most people prefer differentiations over integrations. Therefore we redefine
the representative
\begin{equation}
a({\bf r},t)=\int_{t_o}^t\alpha({\bf r},\tau)
\exp\bigg({\sigma\over\varepsilon}(\tau-t)\bigg)\hbox{d}\tau\ .  \label{ans4}
\end{equation}
This transforms (\ref{ans2}) to
\begin{equation}
{\bf B}({\bf r},t)=\nabla\times{\bf v}
\bigg({\sigma\over\varepsilon}+\partial_t\bigg)a({\bf r},t)      \label{ans2a}
\end{equation}
and (\ref{ans3}) to
\begin{equation}
{\bf E}({\bf r},t)={1\over\varepsilon\mu}
\nabla\times\nabla\times{\bf v}a({\bf r},t)
-{1\over\varepsilon}\int_{t_o}^t {\bf j}_e({\bf r},\tau)
\exp\bigg({\sigma\over\varepsilon}(\tau-t)\bigg)\hbox{d}\tau\ .  \label{ans3a}
\end{equation}
Please find these terms enclosed in the representation formulae
(\ref{rep1}) and (\ref{rep2}).

The only Maxwellian equation (\ref{max2}) not yet fulfilled
produces after insertion of
(\ref{ans2a}) and (\ref{ans3a}) the nonhomogeneous telegraph equation
(\ref{con1}). To see this, one has to apply the lemma of triple curl
(\ref{tricur1}). This is possible only
if the current density is parallel to the supporting field, i.e.\
if condition (\ref{con4}) is fulfilled.

Now that we have taken care of the enforced current ${\bf j}_e({\bf r},t)$,
we may assume for the rest of the proof that only the current caused by
the inner electric field via Ohm's law (\ref{ohm}) remains. This is not
a nonhomogeneity. The first Maxwellian (\ref{max1}) can be simplified
\begin{equation}
\nabla\times{\bf B}({\bf r},t)=\mu\sigma{\bf E}({\bf r},t)
+\varepsilon\mu\partial_t{\bf E}({\bf r},t)\ .                  \label{max1b}
\end{equation}

Third step: The ansatz
\begin{equation}
{\bf E}({\bf r},t)=\nabla\times{\bf v}\beta({\bf r},t)          \label{ans5}
\end{equation}
inserted into the second Maxwellian (\ref{max2}) produces the
representation
\begin{equation}
{\bf B}({\bf r},t)=-\nabla\times\nabla\times{\bf v}
\int_{t_o}^t\beta({\bf r},\tau)\hbox{d}\tau\ .                 \label{ans6}
\end{equation}
Again, for calculational convenience we redefine
\begin{equation}
b({\bf r},t)=\int_{t_o}^t\beta({\bf r},\tau)\hbox{d}\tau       \label{ans7}
\end{equation}
to obtain from (\ref{ans5}) and (\ref{ans6}) the still missing terms
in the general representation formulae (\ref{rep1}) and (\ref{rep2}).
The only Maxwellian which is yet not satisfied, the first one (\ref{max1b}),
causes, after application of the lemma of triple curl (\ref{tricur1}),
the condition (\ref{con2}), again a telegraph equation,
this time, however, a homogeneous one. Q.E.D.

The general representation theorem copes with almost everything discussed
in most monographies on electrodynamics \cite{Jac75}:
electrostatics, magnetostatics,
also electric discarge, skin effects in conductors,
propagation of electromagnetic waves in space, guides and resonating cavities,
optics, metallic optics too, radiation from antennae
and all kinds of electromagnetic scattering, especially Mie scattering,
which appears in thoses monographies as an extremely difficult case.
For the theory of supraconductivity, Ohm's law (\ref{ohm}) must
be replaced with London's equation, but this even simplifies
the derivation of a slightly modified representation theorem.
Only the electrodynamics of nonisotropic materials cannot be tackled
in this way.

Vector potential and double vector potential, dubbed Hertzian vector,
are known since long.
The furthest reaching representation of electromagnetic fields
in terms of these potentials was probably found by Max von Laue
\cite{Lau48}. For instance, the term in (\ref{rep2}) with the integral
over the current density, which bewilders saplings, was given by Laue,
but without consideration of conductivity. However, Laue did not
disentangle Maxwell's equations.
Many scientists subject their vector potentials to so-called Coulomb
or Lorentz gauges. These gauges are related to invariances,
but initial and boundary data break them. Thus in initial- and
boundary-value problems, these vector potentials mislead.
What one must use instead are adaptive scalars times fixed vectors.
One can imagine suitable vector potentials as scalars riding
on prepared vector fields like trains ride on rails.
Inspiring in this direction was Peter Debye's simplified solution
of Mie's problem using a special Hertzian vector \cite{Deb09}.
A remarkably complete list of scalar functions that are useful
for the disentanglement of Maxwell's equations was presented by Meixner
and Sch\"afke \cite{Mei54}. Yet this list is valid only for free
propagation of harmonic electromagnetic waves.
The general principle of representation, i.e.\ the lemma of triple curl,
was obviously unknown to all these scientists. Later on electronic
computers spread and absorbed interest. So this gap in mathematical
physics was filled only in 1985 \cite{Bro85}.

The brightest indication that scientists do not understand the general
principle of representation is the lack of a reasonable theory of
electromagnetic diffraction. There were attempts, for example
\cite{Str39},\cite{Smy47},\cite{Smy69},\cite{The52}, to account
for the vector fields, but they produced after longish explications only
approximations - if at all. Also, partial wave expansions are not
helpful since they converge well only for long waves \cite{Mei54}.
By contrast, the theory that will be developed in the following sections
is straightforward, yields exact equations and is easily applied
to practical problems. Kirchhoff or his scholars had done this
if they had only known the approach.

The situation was similar in hydrodynamics. With
the same methods as explained here it was possible to derive, for
the first time, turbulence in pipes from the Navier-Stokes equation
\cite{Bob88}. In 1989 your author published the prediction
that pipe turbulence consists of transients \cite{Bro89}. It was
verified experimentally in 2006 \cite{Hof06}.

\section{Boundary values on perfect conductors}
\label{sec:boundary}
While the representatives $a$, $b$ and $c$ obey separated differential
equations, they are usually tied together in the conditions on the
boundary- and initial values of ${\bf B}$ and ${\bf E}$. Consider,
for simplicity, the diffraction of electromagnetic waves where
explicit consideration of sources is not necessary. We do not need
a scalar potential $c$. $a$ and $b$ must solve only homogeneous
telegraph equations, cf.\ (\ref{con1}) and (\ref{con2}).

Waves are diffracted when impeded by a screen.
The only way to get on with boundary conditions rather than with
conditions of transition is to have the screen made of perfectly
conducting material. Then, because of Ohm's law
(\ref{ohm}) for $\sigma\rightarrow\infty$, the electric field ${\bf E}$
cannot maintain, on the screen, any tangent component. The waves do not
penetrate.

Let ${\bf t}$ denote any vector tangent to the screen $S$
not to be confounded with the time $t$.
The boundary conditions follow from the representation (\ref{rep2})
\begin{equation}
{1\over\varepsilon\mu}
(\nabla\times\nabla\times{\bf v}a({\bf r},t)){\bf t}
+\partial_t(\nabla\times{\bf v}b({\bf r},t)){\bf t}=0
\hbox{ \ for \ }{\bf r}\ \epsilon\ S\ .                       \label{bou1}
\end{equation}
These are two equations because there are two linearly independent
tangential vectors ${\bf t}$ on a two-dimensional boundary. All the more
it is surprising that the two unknowns $a$ and $b$ can satisfy
the next four equations
\begin{equation}
(\nabla\times\nabla\times{\bf v}a({\bf r},t)){\bf t}=0\hbox{ \ and \ }
(\nabla\times{\bf v}b({\bf r},t)){\bf t}=0
\hbox{ \ for \ } {\bf r}\ \epsilon\ S\ .                      \label{bou2}
\end{equation}
This is possible if the supporting field ${\bf v}$ is
parallel or perpendicular to the diffracting screen.

To prove this, define a local Cartesian coordinate system
\begin{equation}
{\bf r}={\bf e}_x x+{\bf e}_y y+{\bf e}_z z\qquad
{\bf t}={\bf e}_x t_x+{\bf e}_y t_y                            \label{car}
\end{equation}
such that it unit vectors
${\bf e}_x$ and ${\bf e}_y$ be parallel to the screen, whereas ${\bf e}_z$
pierce it normally. Just the components $t_x$ and $t_y$ of the tangential
vector ${\bf t}$ are arbitrary though constant. Consequently only $x$ and $y$
components of curl and double curl need to be considered if the boundary
conditions (\ref{bou2}) are to be satified.

${\bf v}={\bf e}_z$ is according to (\ref{tricur3}) an admissible choice
for the supporting vector field. The equations (\ref{bou2}) become
\begin{eqnarray}
(\nabla\times{\bf e}_z b){\bf t}={\bf e}_x{\bf t}\partial_y b
                              -{\bf e}_y{\bf t}\partial_x b=0\ \nonumber\\
(\nabla\times\nabla\times{\bf e}_z a){\bf t}=
                   {\bf e}_x{\bf t}\partial_x \partial_z a
                  +{\bf e}_y{\bf t}\partial_y \partial_z a=0\,.\label{bou4}
\end{eqnarray}
The differentiations with respect to $x$ and $y$ are inner ones since
the tangent vector ${\bf t}$ of equation (\ref{bou2}) is spanned
by ${\bf e}_x$ and ${\bf e}_y$. Both $\partial_x b=0$ and $\partial_y b=0$
are satisfied on the screen if $b=0$. Equally, both $\partial_x \partial_z a=0$
and $\partial_y \partial_z a=0$ are satisfied on the screen if $\partial_z a=0$.

Generally the supporting vector field ${\bf v}$ (\ref{tricur3}) is
not constant. However, if it is normal to the surface, we can construe
its length as a factor of the representative and repeat the preceding
calculation. Hence the following theorem:

\medskip
\noindent{\bf Theorem on Boundary Conditions.}
{\sl A supporting vector field ${\bf v}$ normal to the surface $S$
of a perfect conductor induces
homogeneous boundary conditions of the first kind for the representative
$b$, whereas the representative $a$ must fulfill homogeneous boundary
conditions of the second kind
\begin{equation}
b({\bf r},t)=0 \hbox{ \ and \ } \partial_n |{\bf v}| a({\bf r},t)=0
\hbox{ \ for \ } {\bf r}\ \epsilon\ S\ .                    \label{bou5}
\end{equation}
Oppositely, a supporting vector field tangent to the surface induces
homogeneous boundary conditions of the first kind for the representative
$a$, whereas the representative $b$ must fulfill homogeneous boundary
conditions of the second kind
\begin{equation}
a({\bf r},t)=0 \hbox{ \ and \ } \partial_n b({\bf r},t)=0
\hbox{ \ for \ } {\bf r}\ \epsilon\ S\ .                    \label{bou8}
\end{equation}
$\partial_n$ denotes differentiation along the normal on $S$.}
\medskip

For convenience of reference, your author bundled the essentials
of this section in one theorem. Its second part still has to be proven.
With the local coordinate system introduced above,
${\bf v}={\bf e}_x$ is according to (\ref{tricur3}) an admissible choice
for the supporting vector field, too. The equations (\ref{bou2}) are now
\begin{eqnarray}
(\nabla\times{\bf e}_x b){\bf t}={\bf e}_y{\bf t}\partial_z b=0\ \nonumber\\
(\nabla\times\nabla\times{\bf e}_x a){\bf t}=
{\bf e}_x{\bf t}(\partial_x^2-\varepsilon\mu\partial_t^2-\mu\sigma\partial_t)a
+{\bf e}_y{\bf t}\partial_x \partial_y a=0\,.                    \label{bou7}
\end{eqnarray}
For the first component of the latter equation the homogenous telegraph
equation (\ref{con1}) was exerted. Again,
the differentiations with respect to $x$ and $y$ are inner ones. The
same is true for the differentiations with respect to $t$ because
boundary conditions must hold for all times. Hence $\partial_z b=0$
and $a=0$ on the screen. When this is written without Cartesian
coordinates, the second part of the above theorem emerges. A correction
with the length of the supporting vector is not necessary here because
${\bf v}$ as defined in (\ref{tricur3}) does not vary in the direction
of the normal if it is perpendicular to that normal.
Q.E.D.

Screens do not enclose radiation. Much of it spreads in open space.
Thus the boundary conditions need to be completed, namely by retardation
\begin{equation}
a\hbox{ or }b({\bf r},t)\sim\bigg(
{f_{a\hbox{ or }b}(|{\bf r}|-t/\sqrt{\varepsilon\mu})\over|{\bf r}|}+
O(\sigma)\bigg)\,e^{-\sigma t/2\varepsilon}
\hbox{ \ for \ }|{\bf r}|\rightarrow\infty                  \label{bou10}
\end{equation}
meaning that waves trail away in nirvana and never return.
The functions $f_a$ and $f_b$ may depend on the direction of the radiation,
but they depend on the distance $|{\bf r}|$ only through the
compound argument $|{\bf r}|-t/\sqrt{\varepsilon\mu}$.
$O(\sigma)$ is E.Landau's order symbol to appraise neglected terms
on the right-hand side, see e.g.\ \cite[Section~1.1]{Erd56}.
Notice: $\sigma$ denotes here the conductivity of the propagating
medium, for example air.

Hence we may have boundary conditions that do not couple
the representatives. Yet to profit from the theorem we need
supporting vector fields that are
{\sl everywhere\/} either normal or tangent to the screen.
According to (\ref{tricur3}) this can be achieved
for four types of screens: for plane ones, for parts of
spheres, for parts of cones and for parts of cylinders, i.e.\ for
cylinders with arbitrary cross sections.
In this article we will be busy enough
to cope with diffraction by plane screens and will use the theorem
with the tangent supporting vector field (\ref{bou8}).

\section{Initial values transformed}
\label{sec:initial}
We account for arbitrary initial values using a Fourier or rather
a Laplace transform. All fields are proportional to $\exp(-i\omega t)$.
\begin{eqnarray}
&{\bf B}({\bf r},t)={\bf B}_k({\bf r})\ e^{-i\omega t}\qquad
&{\bf E}({\bf r},t)={\bf E}_k({\bf r})\ e^{-i\omega t}          \label{ini1}\\
&a({\bf r},t)=a_k({\bf r})\ e^{-i\omega t}\qquad
&b({\bf r},t)=b_k({\bf r})\ e^{-i\omega t}                      \label{ini2}
\end{eqnarray}
The telegraph equations (\ref{con1}-\ref{con2})
become thus Helmholtz equations
\begin{equation}
\nabla^2 a_k({\bf r})+k^2 a_k({\bf r})=0\qquad
\nabla^2 b_k({\bf r})+k^2 b_k({\bf r})=0\ .                     \label{hel2}
\end{equation}
The wave number $k$ depends on the frequency $\omega$ according to
\begin{equation}
k^2=\varepsilon\mu\omega^2+i\mu\sigma\omega\ .                  \label{kom1}
\end{equation}
One of the two parameters, $\omega$ or $k$, can be chosen as real.
In the latter case, a negative imaginary part of $\omega$ describes
fading with time $t\rightarrow\infty$. This is expressed through
\begin{equation}
\omega={k\over\sqrt{\varepsilon\mu}}-i{\sigma\over 2\varepsilon}+O(\sigma^2)
                                                                \label{kom2a}
\end{equation}
and corresponds to equation (\ref{bou10}). In the first case,
a positive imaginary part of $k$ describes attenuation
in space as $|{\bf r}|\rightarrow\infty$.
Modulus and phase of $k$ can be read from
\begin{equation}
k=\root 4 \of{\varepsilon^2\mu^2\omega^4+\mu^2\sigma^2\omega^2}\
\exp\bigg({i\over 2}\arctan{\sigma\over\varepsilon\omega}\bigg)\ .\label{kom2}
\end{equation}
The point to be made is that the phase of $k$ varies only between 0 and
$\pi/4$. This will matter in the discussions of section \ref{sec:error}.

Finally we must translate retardation (\ref{bou10}) into the language of
Fourier transforms. Both representatives $a_k$ and $b_k$
behave as leaving spherical waves
\begin{equation}
a\hbox{ or }b_k({\bf r})=F_{a\hbox{ or }b}{\exp(ik|{\bf r}|)\over|{\bf r}|}
+O(|{\bf r}|^{-2})\hbox{ \ for \ }|{\bf r}|\rightarrow\infty \label{ini5}
\end{equation}
at which the real parts of $k$ and $\omega$ are supposed to carry
the same sign. The scattering amplitudes $F_a$ and $F_b$
may depend on the wave number $k$ and the direction of the radiation,
but not on $|{\bf r}|$. Though Sommerfeld preferred to write this as
$\partial_{|{\bf r}|}a_k({\bf r})=ika_k({\bf r})+O(|{\bf r}|^{-2})$ etc.,
your author persists in calling (\ref{ini5}) Sommerfeld's
radiation condition.

\section{Kirchhoff's theory corrected}
\label{sec:kirchhoff}
Altogether we found that, for a plane screen, the representatives
$a_k({\bf r})$ and $b_k({\bf r})$ can be computed separately. Both satisfy
Helmholtz equations (\ref{hel2}) and exhibit the same behavior in infinity
(\ref{ini5}). Yet on the screen $a_k({\bf r})$ must solve a boundary-value
problem of the first kind, whereas $b_k({\bf r})$ is subject to
boundary conditions of the second kind, see equations (\ref{bou8}).
\footnote{From now on, the
index $p$ at the position of the probe ${\bf r}_p$ is indispensable again
as there are other locations which must be discriminated, viz.\
the position of the source ${\bf r}_q$ and arbitrary points on the
screen ${\bf r}$.}

Modifications of Kirchhoff's integral (\ref{kir1}) inaugurated by
Sommerfeld are useful to solve both boundary-value problems. Instead
of the spherical wave $\exp(ik|{\bf r}_p-{\bf r}|)/|{\bf r}_p-{\bf r}|$
flexible Green functions $G({\bf r}_p,{\bf r})$
\begin{equation}
\psi({\bf r}_p)={1\over 4\pi}\int\!\!\int_F\bigg(
G({\bf r}_p,{\bf r})\ \partial_n\psi({\bf r})
-\psi({\bf r})\ \partial_n G({\bf r}_p,{\bf r})
\bigg)\ \hbox{d}f                                                \label{kir2}
\end{equation}
are introduced. Kirchhoff's formula remains valid
if these Green functions fulfill the Helmholtz equation (\ref{hel1}),
if they have the same singularity as the
spherical wave
\begin{equation}
\partial_{|{\bf r}_p-{\bf r}|}G({\bf r}_p,{\bf r})=
-({\bf r}_p-{\bf r})^{-2}+O(|{\bf r}_p-{\bf r}|^{-1})
\hbox{ for }{\bf r}_p\rightarrow {\bf r}                         \label{kir3}
\end{equation}
and respect the radiation condition as in equation (\ref{ini5}).

In a boundary-value problem of the first kind, the values of the function
$\psi({\bf r})$ itself are known on the plane, but no information
on the values of the derivative $\partial_n\psi({\bf r})$ is available
beforehand. Hence we need a Green function that is zero on the plane
lest the unknown values matter. We find from (\ref{kir2})
\begin{equation}
\psi({\bf r}_p)={-1\over 4\pi}\int\!\!\int_F
\partial_n G_1({\bf r}_p,{\bf r})\ \psi({\bf r})\,\hbox{d}f
\hbox{ \ if \ }G_1({\bf r}_p,{\bf r})=0
\hbox{ \ for \ }{\bf r}\ \epsilon\ F\ .                        \label{kir5}
\end{equation}
On the other hand, if the values of the derivative
are known, we must do without the function itself.
For the solution of the boundary-value problem
of the second kind we need a second Green function
\begin{equation}
\psi({\bf r}_p)={1\over 4\pi}\int\!\!\int_F
G_2({\bf r}_p,{\bf r})\ \partial_n \psi({\bf r})\,\hbox{d}f
\hbox{ \ if \ }\partial_n G_2({\bf r}_p,{\bf r})=0
\hbox{ \ for \ }{\bf r}\ \epsilon\ F\ .                        \label{kir6}
\end{equation}
For $F$ being a plane characterized in Cartesian coordinates (\ref{car})
by, say, z=0, both Green functions are easily found as a spherical wave
around the point of measurement ${\bf r_p}$ plus its mirrored image on
the other side of the plane
\begin{equation}
G_{1\hbox{ or }2}({\bf r}_p,{\bf r})=
{\exp(ik|{\bf r}_p-{\bf r}|)\over|{\bf r}_p-{\bf r}|}-\hbox{or}+
{\exp(ik|{\bf r}_m-{\bf r}|)\over|{\bf r}_m-{\bf r}|}       \label{kir8}
\end{equation}
with
\begin{equation}
{\bf r}_p={\bf e}_x x_p+{\bf e}_y y_p+{\bf e}_z z_p \qquad
{\bf r}_m={\bf e}_x x_p+{\bf e}_y y_p-{\bf e}_z z_p\ .       \label{kir9}
\end{equation}
Both functions own the correct singularity as in (\ref{kir3}) and respect
the radiation condition (\ref{ini5}).

Furthermore it shall be assumed that the primary wave is created
by a point-like source at ${\bf r_q}$ behind the plane
\begin{equation}
\psi({\bf r})=
{\exp(ik|{\bf r}-{\bf r}_q|)\over|{\bf r}-{\bf r}_q|}\ .   \label{kir10}
\end{equation}

This kind of primary wave is to be preferred over the usual plane wave
because it respects a radiation condition of type (\ref{ini5}) and
ensures thus the validity of the generalized Kirchhoff formula (\ref{kir2}).
Also it is closer to experiments as it is simpler to produce an
approximate spherical wave than an approximate plain wave.
The beloved plain wave can be obtained from the spherical wave
by a straightforward limiting transition $|{\bf r}_q|\rightarrow\infty$.
Radiation from higher multipoles can be derived by differentiation
with respect to ${\bf r}_q$ and the general primary wave by superposition.

Writing down the integral for the boundary conditions of the first kind
(\ref{kir5}) in Cartesian coordinates ends up with
\begin{equation}
\psi({\bf r}_p)={\exp(ikR_{pq})\over R_{pq}}\ {-z_pR_{pq}\over 2\pi}
\int_{-\infty}^{+\infty}\int_{-\infty}^{+\infty}{ik-1/R_p(x,y)\over R_p^2(x,y)R_q(x,y)}\
e^{ik\Delta(x,y)}\hbox{d}x\hbox{d}y                               \label{kir11}
\end{equation}
and the integral for the boundary conditions of the second kind (\ref{kir6})
becomes
\begin{equation}
\psi({\bf r}_p)={\exp(ikR_{pq})\over R_{pq}}\ { z_qR_{pq}\over 2\pi}
\int_{-\infty}^{+\infty}\int_{-\infty}^{+\infty}{ik-1/R_q(x,y)\over R_q^2(x,y)R_p(x,y)}\
e^{ik\Delta(x,y)}\hbox{d}x\hbox{d}y                               \label{kir12}
\end{equation}
with
\begin{eqnarray}
R_p(x,y)=|{\bf r}_p-{\bf r}(x,y)|=\sqrt{(x-x_p)^2+(y-y_p)^2+z_p^2}\nonumber\\
R_q(x,y)=|{\bf r}(x,y)-{\bf r}_q|=\sqrt{(x-x_q)^2+(y-y_q)^2+z_q^2}
                                                                 \label{tri3}\\
R_{pq}=|{\bf r}_p-{\bf r}_q|=\sqrt{(x_p-x_q)^2+(y_p-y_q)^2+(z_p-z_q)^2}\nonumber
\end{eqnarray}
and the {\sl triangle function\/}
\begin{equation}
\Delta(x,y)=R_p(x,y)+R_q(x,y)-R_{pq}                              \label{tri1}
\end{equation}
which measures the difference of lengths when the points ${\bf r}_p$
and ${\bf r}_q$ are either directy connected
or via an arbitrary point ${\bf r}={\bf e}_x x+{\bf e}_y y$ on the
intermediate plane.

It is not by accident that the $\psi$'s in the equations (\ref{kir11})
and (\ref{kir12}) are identical. The factors behind the spherical wave
$\exp(ikR_{pq})/R_{pq}$ including the double integrals extending from
$-\infty$ to $+\infty$ have both the value 1; a proof of this fact
will be given in section \ref{sec:imaging}. The formulae (\ref{kir11})
and (\ref{kir12}) are just different
mathematical realizations of Huygens' principle:
A wave propagating from a source point that strikes a plane excites
there secondary waves interfering to reproduce the original wave.

Diffraction happens only if a screen covers parts of the plane.
Using the boundary conditions in the theorem with the tangent
supporting field (\ref{bou8}) yields explicit expressions
for the representatives
\begin{equation}
a_k({\bf r}_p)={\exp(ikR_{pq})\over R_{pq}} I\!I({\bf r}_p,{\bf r}_q)
\qquad
b_k({\bf r}_p)={\exp(ikR_{pq})\over R_{pq}} I\!I({\bf r}_q,{\bf r}_p)
                                                                  \label{kir13}
\end{equation}
with the double integral
\begin{equation}
I\!I({\bf r}_p,{\bf r}_q)={-|z_p|R_{pq}\over 2\pi}\int\!\!\int_D
{ik-1/R_p(\xi,\eta)\over R_p^2(\xi,\eta)R_q(\xi,\eta)}\
{\partial(x,y)\over\partial(\xi,\eta)} \
e^{ik\Delta(\xi,\eta)}\hbox{d}\xi\hbox{d}\eta\ .                  \label{kir14}
\end{equation}
To be general enough for all applications, arbitrary variables $\xi$ and
$\eta$ shall substitute the Cartesian ones $x(\xi,\eta)$ and $y(\xi,\eta)$.
The functional determinant $\partial(x,y)/\partial(\xi,\eta)$ enters
to transform the element of the surface
$\hbox{d}f=\hbox{d}x\hbox{d}y$. The integral $I\!I$ extends only
over the aperture or diaphragm $D$ in the surface $F$. The arbitrariness
of $\xi$ and $\eta$ shall be used for a simple description of the
aperture in the way that freely varying $\xi$ with a fixed $\eta$
depicts an edge. The aperture is described, for example, by
$-\infty<\xi<\infty$ and $-\eta_-<\eta<\eta_+$.

The enthusing result is that, for the full electromagnetic theory
of diffraction with all possible polarizations,
one needs to evaluate only one integral. There is scarcely
more work to be done than in the theory for scalar waves.
The addititional work consists in some elementary
differentiations as prescribed in the representation formulae
(\ref{rep1}-\ref{rep2}) simplified for Fourier transforms as
\begin{eqnarray}
{\bf B}_k({\bf r}_p)=
\bigg({\sigma\over\varepsilon}-i\omega\bigg)
\nabla_p\times{\bf t} a_k({\bf r}_p)
-\nabla_p\times\nabla_p\times{\bf t} b_k({\bf r}_p)         \label{kir15}\\
{\bf E}_k({\bf r}_p)={1\over\varepsilon\mu}
\nabla_p\times\nabla_p\times{\bf t} a_k({\bf r}_p)
-i\omega\nabla_p\times{\bf t} b_k({\bf r}_p)                \label{kir16}
\end{eqnarray}
with a supporting vector field {\bf v}={\bf t}
that consists of a constant tangent as in (\ref{car}).

The equations (\ref{kir13}) through (\ref{kir16}) constitute
the second major item of this article.
The integral (\ref{kir14}) can be evaluated using various techniques.
Your author will discuss in the following sections only one:
the method of stationary phase applicable for short waves with
$\Re k\rightarrow\infty$ and $\Im k>0$ as conditions on
the real and imaginary parts of the wave number $k$, respectively.

While equations (\ref{kir13}-\ref{kir14}) solve mathematically
well-posed boundary-value problems without any error, they do not describe
physical diffraction exactly. When a conducting screen diffracts a wave,
reflection cannot be avoided. Most of the
reflected wave stays in outer space, $z_p<0$, but it is
also diffracted. A tiny fraction of reflection invades
inner space $z_p>0$. We will learn to handle this in section
\ref{sec:imaging}.

\section{Light and shadow}
\label{sec:triangle}
The triangle function rules the diffraction of short waves.
In its definition (\ref{tri1}) interest was focused
on the point of the screen ${\bf r}$.
Yet the triangle function also depends on the points of
probe ${\bf r}_p$ and source ${\bf r}_q$
\begin{equation}
\Delta({\bf r},{\bf r}_p,{\bf r}_q)=
|{\bf r}_p-{\bf r}|+|{\bf r}-{\bf r}_q|-|{\bf r}_p-{\bf r}_q|\ . \label{lig1}
\end{equation}
Your author decided to locate
\begin{eqnarray}
\hbox{screen }{\bf r}=&{\bf e}_x x+{\bf e}_y y+{\bf e}_z z
                                           &\hbox{ \ at }z=0\ , \nonumber\\
\hbox{probe }{\bf r}_p=&{\bf e}_x x_p+{\bf e}_y y_p+{\bf e}_z z_p
                                         &\hbox{ \ at }z_p>0\ ,\label{loc}\\
\hbox{source }{\bf r}_q=&{\bf e}_x x_q+{\bf e}_y y_q+{\bf e}_z z_q
                                         &\hbox{ \ at }z_q<0\ .\nonumber
\end{eqnarray}

The triangle function is positive except at that point
${\bf r}={\bf r}_s={\bf e}_x x_s+{\bf e}_y y_s$
where the straight line between probe and source pierces the plane
\begin{equation}
x_s+iy_s={z_p(x_q+iy_q)-z_q(x_p+iy_p)\over z_p-z_q}\ .          \label{tri4}
\end{equation}
Complex notation is preferred because it eases transformation to
other coordinate systems, see below.
The point ${\bf r}_s$ is the location of the absolute minimum
of the triangle function, and all the more it is a {\sl stationary point\/}.

The consideration holds for fixed points of source and probe.
However, if only ${\bf r}_q$ is fixed, whereas ${\bf r}_p$ varies
while ${\bf r}={\bf r}_s$ slides on the edge of the screen, then
\begin{equation}
\Delta({\bf r},{\bf r}_p,{\bf r}_q)=0
          \hbox{ \ if }{\bf r}\hbox{ on the edge}               \label{lig2}
\end{equation}
determines as function of ${\bf r}_p$ a surface, namely the {\sl border
of shadow\/}.

For calculating diffraction, we will need the root of the
triangle function. Your author utilizes the ambiguity of the root
to demand
\begin{equation}
\sqrt{\Delta({\bf r},{\bf r}_p,{\bf r}_q)}=
\cases{+|\sqrt{\Delta({\bf r},{\bf r}_p,{\bf r}_q)}|
                              &if ${\bf r}_p$ in the shadow,\cr
       -|\sqrt{\Delta({\bf r},{\bf r}_p,{\bf r}_q)}|
                              &if ${\bf r}_p$ in the light.\cr} \label{lig3}
\end{equation}
It is cogent to assign different signs to the dark and the bright
if the function is to be differentiable. The triangle function
is an analytic function which depends, around its minimum
defined by (\ref{lig2}), quadratically on its variables.
Therefore omitting the signs in (\ref{lig3}) would
induce a similar discontinuity as in the assignment $\sqrt{x^2}=|x|$.
The absolute assignment of the sign, on the contrary, is arbitrary since
diffraction either by a screen or its complement is equal; remember
Babinet's principle \cite[\S~47]{Bor65}.

To decide where there is light or shadow,
a handy criterion is needed. There is light on the probe if the screen
does not impede the straight connection between source ${\bf r}_q$
and probe ${\bf r}_p$. Thus the positive sign in (\ref{lig3})
is to be taken if the stationary point $x_s+iy_s$ in (\ref{tri4})
misses the aperture.

\medskip
\noindent {\bf Criterion of Light.} {\sl
When one uses transformed coordinates $\xi,\eta$ adapted to the screen
such that the aperture is described by $\eta_-<\eta<\eta_+$
while $\xi$ varies freely, the negative sign of triangle function's
root (\ref{lig3}) has to be taken if
\begin{equation}
\eta_-<\eta_s<\eta_+\ .                                  \label{lig4}
\end{equation}
$\eta_s$ is calculated from the stationary point (\ref{tri4})
via coordinate transformation.
}
\medskip

For example on a screen with a circular aperture,
cylindrical coordinates $\rho,\varphi,z=0$ suit.
The aperture is defined by $\rho<\rho_0$ with $\rho_0$ as the radius
of the stop, while $\varphi$ varies freely.
The transformation to cylindrical coordinates is facilitated through
\begin{equation}
x_s+iy_s=\rho_s e^{i\varphi_s}\qquad
x_p+iy_p=\rho_p e^{i\varphi_p}\qquad
x_q+iy_q=\rho_q e^{i\varphi_q}\qquad                              \label{tri5}
\end{equation}
with the $\rho$'s as axial distances and the $\varphi$'s as azimuthal angles.
Insertion into (\ref{tri4}) produces two equivalent formulas for the
axial distance
\begin{eqnarray}
\rho_s={\sqrt{(z_p\rho_q-z_q\rho_p)^2+4z_pz_q\rho_p\rho_qS^2}\over z_p-z_q}
                                                               \nonumber\\
      ={\sqrt{(z_p\rho_q+z_q\rho_p)^2-4z_pz_q\rho_p\rho_qC^2}\over z_p-z_q}
                                                               \label{tri7}
\end{eqnarray}
with the abbreviations
\begin{equation}
C=\cos{\varphi_p-\varphi_q\over 2}\qquad
S=\sin{\varphi_p-\varphi_q\over 2}\ .                          \label{tri8}
\end{equation}
$\rho_s>\rho_0$ is thus the criterion for the domain of shadow,
i.e.\ for the positive sign in (\ref{lig3}).

Instead of applying elementary geometry, as was done in this section,
one may calculate the stationary point $\xi_s,\eta_s$ by simultaneous
solution of the equations
\begin{equation}
\partial_\xi\Delta(\xi,\eta)=0\qquad\partial_\eta\Delta(\xi,\eta)=0\qquad
\leftrightarrow\qquad \xi=\xi_s\qquad\eta=\eta_s\ .          \label{tri14}
\end{equation}
The result is, of course, the same as that given in (\ref{tri4}) with
subsequent transformation of coordinates, but the computational effort
is larger. Your author displays the equations (\ref{tri14}) only
to ease comprehension of the astonishing equation (\ref{exh1})
which will appear in section \ref{sec:exhaust}.

\section{Using the error function for diffraction}
\label{sec:error}
In the theory of diffraction, Fresnel integrals
\begin{equation}
C_1(z)=\sqrt{{2\over\pi}}\int_o^z \cos w^2\ \hbox{d}w\qquad
S_1(z)=\sqrt{{2\over\pi}}\int_o^z \sin w^2\ \hbox{d}w           \label{err1}
\end{equation}
are still custom, but the error function or rather the {\sl complementary
error function\/} is handier \cite{Abr70},\cite{Jah66}
\begin{equation}
\erfc(z)={2\over\sqrt{\pi}}\int_z^\infty e^{-w^2}\hbox{d}w\ .   \label{err3}
\end{equation}
All statisticians become perfect opticians when they are willing
to handle their favorite function with complex argument.

The error function comprises the Fresnel integrals in a similar fashion
as the exponential function contains sine and cosine
\begin{equation}
\erfc(\sqrt{-i}z)=1-\sqrt{-i}\sqrt{2}\ \big((C_1(z)+iS_1(z)\big)\ .\label{err4}
\end{equation}
If $z$ is assumed as real, $\sqrt{-i}z$ varies on the
second main diagonal of the complex plane since
$\sqrt{-i}=(1-i)/\sqrt{2}=\exp(-i\pi/4)$.
The features known for real argument remain if the complex argument
is enclosed between the first and the second main diagonals of the
complex plane; expressed by a relation between imaginary and real parts:
$|\Im z|\le |\Re z|$. This is exactly what we need for optics,
see (\ref{kom2}).
For large negative real parts $\Re z\rightarrow-\infty$
of the argument the complementary error function starts
at the value 2, assumes the value 1 at the origin $z=0$, and attains
the value 0 for large positive real parts $\Re z\rightarrow+\infty$.
In the crudest approximation, one may think
of the complementary error function as 2 for negative arguments and 0
for positive ones. It is a switch.

The asymptotic expansion of the error function familiar
on the real axis remains valid in the wedge between the main diagonals
which contains the real axis
\begin{equation}
\erfc(z)={1\over \sqrt{\pi}z}\ e^{-z^2}(1+O(|z|^{-2})
\hbox{ \ for \ }\Re z \rightarrow+\infty\ .                      \label{err5}
\end{equation}
The asymptotic expansion on the other side of the complex plane
$\Re z<0$ follows from
\begin{equation}
\erfc(z)=2-\erfc(-z)\ .                                          \label{err6}
\end{equation}

The only refinement due to complexity is that the complementary error
function decreases monotonously on the real axis, whereas it takes complex
values and both real and imaginary parts oscillate when the argument
becomes complex.

\section{The method of stationary phase for two-dimensional integrals}
\label{sec:exhaust}
Let us recall the asymptotic calculation of one-dimensional integrals
\begin{equation}
I(k,\eta_-,\eta_+)=\int_{\eta_-}^{\eta_+} A(\eta)\
           e^{ik\Delta(\eta)}\ d\eta                            \label{stat1}
\end{equation}
for $\Re k\rightarrow\infty$ with $\Im k>0$.
We assume that the real function
$\Delta(\eta)\ge 0$ is stationary for some $\eta=\eta_s$, i.e.
\begin{equation}
\Delta(\eta)={\Delta_{\eta\eta}\over 2}\ (\eta-\eta_s)^2 +
             O((\eta-\eta_s)^3)\ .                             \label{stat2}
\end{equation}
The indices at $\Delta$ indicate that the function be differentiated
twice and the result evaluated at $\eta=\eta_s$.

The familiar approach in the method of stationary phase is
to introduce a new variable
\begin{equation}
 v= \sqrt{{\Delta_{\eta\eta}\over 2}}\ (\eta-\eta_s)\         \label{stat3}
\end{equation}
and to forget the higher-order terms $O((\eta-\eta_s)^3)$ in equation
(\ref{stat2}). The function
\begin{equation}
\delta(v)\sim \sqrt{{k \over \pi i}}\ e^{ikv^2}
\hbox{ \ for \ } \Re k\rightarrow\infty,\ \Im k>0             \label{stat4}
\end{equation}
may be construed as a representation of Dirac's delta function.
Thus the integral (\ref{stat1}) yields
\begin{equation}
I(k,-\infty,+\infty)\sim \sqrt{{2\pi i\over k\Delta_{\eta\eta}}}\
     A(\eta_s)\ .                                             \label{stat5}
\end{equation}
The amplitude $A(\eta)$ appears as a constant.

If the limits of integration $\eta_\pm$ are $\pm\infty$, this is correct,
but for finite limits the local approximation (\ref{stat3}) induces
systematic errors. What we have to use instead is the {\it global map\/}
$\eta\rightarrow v$
\begin{equation}
v=\sqrt{\Delta(\eta)}\ .                                       \label{stat6}
\end{equation}
To rewrite the integral from the variable $\eta$ to the variable $v$,
we must calculate the differential $d\eta=(d\eta/dv)dv=(dv/d\eta)^{-1}dv$.
Because of (\ref{stat4}) it is sufficient to know the value
of the differential for $v=0$ corresponding to $\eta=\eta_s$.
Thus the value of the differential of the global map (\ref{stat6})
to be used in the integral is the same as the differential
of the local approximation (\ref{stat3}).
The peculiarity of the global map appears only in the limits:
\begin{equation}
I(k,\eta_-,\eta_+)\sim \sqrt{{2\pi i\over k\Delta_{\eta\eta}}}\ A(\eta_s)
        {\erfc\sqrt{-ik\Delta(\eta_-)} -
         \erfc\sqrt{-ik\Delta(\eta_+)}\over 2}\ .              \label{stat7}
\end{equation}

While the preceding is not familiar, it is known \cite[Section~2.9]{Erd56}.
Yet in the theory of diffraction one needs to evaluate two-dimensional
integrals
\begin{equation}
I\!I(k,\eta_-,\eta_+)=\int_{\eta_-}^{\eta_+} \int_{-\infty}^\infty
        A(\xi,\eta)\ e^{ik\Delta(\xi,\eta)}\ d\xi d\eta\ .    \label{stat8}
\end{equation}
Again it is assumed that the real function $\Delta(\xi,\eta)\ge 0$
is stationary at $\xi_s,\eta_s$
\begin{eqnarray}
\Delta(\xi,\eta)=
{\Delta_{\xi\xi}\over 2}\ (\xi-\xi_s)^2 +
\Delta_{\xi\eta}\ (\xi-\xi_s)(\eta-\eta_s) +
{\Delta_{\eta\eta}\over 2}\ (\eta-\eta_s)^2                 \nonumber\\
           + O(|\xi-\xi_s|^3+(|\eta-\eta_s|^3)\ .           \label{stat9}
\end{eqnarray}
It seems to be a hitherto unsolved problem to find a suitable
two-dimensional global map
$\xi,\eta\rightarrow u,v$ such that
\begin{equation}
u^2+v^2=\Delta(\xi,\eta)\ .                              \label{stat10}
\end{equation}
Here is the solution: The map is
\begin{eqnarray}
u=\sqrt{\Delta(\xi,\eta)-\Delta_s(\eta)}\                    \label{exh3}\\
v=\sqrt{\Delta_s(\eta)}\,.                                   \label{exh4}
\end{eqnarray}
The function $\Delta_s(\eta)$ is determined by the

\medskip
\noindent{\bf Principle of Utter Exhaust.}
{\sl Eliminate $\xi$ from the derivative
\begin{equation}
\partial_\xi\Delta(\xi,\eta)=0\ \leftrightarrow\ \xi=\Xi_s(\eta)
                                                           \label{exh1}
\end{equation}
to find the {\it exhausting dependence\/} $\xi=\Xi_s(\eta)$. Insert the
exhausting dependence into the function $\Delta(\xi,\eta)$
to obtain the {\it exhausting function\/}
\begin{equation}
\Delta_s({\eta})=\Delta(\Xi_s(\eta),\eta)\ .                  \label{exh2}
\end{equation}
The integral (\ref{stat8}) is, for $\Re k\rightarrow\infty$ and $\Im k>0$,
asymptotically equal to
\begin{equation}
I\!I(k,\eta_-,\eta_+)\sim {2\pi i\ A(\xi_s,\eta_s) \over
 k \sqrt{\Delta_{\xi\xi}\Delta_{\eta\eta}-\Delta_{\xi\eta}^2}}
        {\erfc\sqrt{-ik\Delta_s(\eta_-)}-
         \erfc\sqrt{-ik\Delta_s(\eta_+)}\over 2}\ .         \label{exh5}
\end{equation}
}
\medskip

Utter exhaust follows from the following indispensable requirements:
\begin{eqnarray}
\xi=\xi_s,\eta=\eta_s \hbox{ be mapped to\ } u=0,v=0\ ,     \label{stat11}\\
u,v \hbox{ be real for all\ } \xi,\eta\ ,                   \label{stat12}\\
v^2=f(\eta) \hbox{ be a function of $\eta$ only}\ .         \label{stat13}
\end{eqnarray}
We need the third requirement (\ref{stat13}) to retain the simplicity
of the limits in the integral $I\!I$ (\ref{stat8}) when mapping
$\xi,\eta$ to $u,v$. Because of equation (\ref{stat10})
and the requirement (\ref{stat12}) the function $f(\eta)$
must never exceed $\Delta(\xi,\eta)$ whatever value $\xi$ takes.
Nevertheless for every $\eta$ there must exist $\xi=\Xi_s(\eta)$
such that equality is reached: $f(\eta)=\Delta(\Xi_s(\eta),\eta)$.
Otherwise $u=\sqrt{\Delta(\xi,\eta)-f(\eta)}$ cannot take 0,
a contradiction to the requirement (\ref{stat11}). In other words,
$f(\eta)$ and $\Delta(\xi,\eta)$ coincide at that value of
$\xi=\Xi_s(\eta)$ where $\Delta(\xi,\eta)$ gets stationary
with respect to $\xi$. Hence $\Xi_s(\eta)$ is determined
by elimination of $\xi$ in the condition (\ref{exh1}).

The final formula (\ref{exh5}) can be understood when compared
to formula (\ref{stat7}). The double integration generates the
factor $\pi i/k$, the square of $\sqrt{\pi i/k}$. The determinant in
$2/\sqrt{\Delta_{\xi\xi}\Delta_{\eta\eta}-\Delta_{\xi\eta}^2}$
of the map $\xi,\eta\rightarrow u,v$ replaces the differential in
$\sqrt{2/\Delta_{\eta\eta}}$ of the map $\eta\rightarrow v$. Q.E.D.

The principle was named as of utter exhaust because the function
$\Delta_s(\eta)$ is the largest possible function of $\eta$ only that
fulfills $\Delta_s(\eta)\le \Delta(\xi,\eta)$; it takes from the two-variable
function as much as a one-variable function can afford, cf.\ equation
(\ref{exh3}).

A further intuitive interpretation follows from a comparison of
equation (\ref{exh1}) with the equations (\ref{tri14}). The stationary
point is where the function $\Delta(\xi,\eta)$ takes its absolute extremum.
On the exhausting dependence, by contrast, $\Delta(\xi,\eta)$ is
extremized only with respect to the one variable $\xi$, whereas the
other variable $\eta$ is fixed. In optics, $\Delta(\xi,\eta)$ essentially
measures the distance between points. The absolutely shortest connection
is of course a straight line. But the exhausting function $\Delta_s(\eta)$
measures a conditionally shortest distance, namely if the connecting line
is forced to touch the edge of the diffracting screen. One can determine
the exhausting function experimentally using a ribbon of rubber and
a lubricated model of the edge.

The difficult part of utter exhaust is the elimination according
to equation (\ref{exh1}). It is therefore gratifying to possess
linear approximations of equations (\ref{exh3}-\ref{exh4}),
similar to the one-dimensional case (\ref{stat3}). These
approximations are
\begin{eqnarray}
u\approx\sqrt{{\Delta_{\xi\xi}\over 2}}\ (\xi-\xi_s)+
 {\Delta_{\xi\eta}\over \sqrt{2 \Delta_{\xi\xi}}}\ (\eta-\eta_s) \label{exh7}\\
v\approx \sqrt{{\Delta_{\xi\xi}\Delta_{\eta\eta}
-\Delta_{\xi\eta}^2\over 2 \Delta_{\xi\xi}}}\ (\eta-\eta_s)\ .   \label{exh8}
\end{eqnarray}
They were found by utter exhaust (\ref{exh1}) applied
to the quadratic terms on the right-hand side of (\ref{stat9});
searching principal axes of the ellipse is not a good idea.
For the integral $I\!I$ (\ref{stat8}), equation (\ref{exh5})
can be used when the exhausting function is approximated as
\begin{equation}
\Delta_s(\eta_{+\hbox{or}-})\approx {\Delta_{\xi\xi}\Delta_{\eta\eta}
-\Delta_{\xi\eta}^2\over 2 \Delta_{\xi\xi}}
\ (\eta_{+\hbox{or}-} -\eta_s)^2\ .                             \label{exh9}
\end{equation}
The relation to the border of shadow discussed in
section \ref{sec:triangle} appears here at first sight.

The method of stationary phase is sometimes blamed as not being
mathematical stringent. It is argued that certain integrals do not
converge and thus certain errors cannot be estimated. The criticism
does not apply here. Guided by mother nature, we made a theory where the
parameter $k$ of asymptoticity has a positive imaginary part
$\Im k>0$. This guarantees the convergence of those integrals. We can
even write the complete asymptotic series.

Augustin Fresnel invented
the {\sl zone construction\/} in 1816 to prove that light travels
as a wave, but this was just a semi-quantitive idea \cite[p.247]{Fre66},
\cite[\S~8.2]{Bor97}. Its mathematical solution is the principle
of utter exhaust presented only now.

\section{The universal formula of diffraction}
\label{sec:universal}
To make use of stationary phase for diffraction,
the amplitude in the integral (\ref{kir14})
\begin{equation}
A(\xi,\eta)\sim
{-z_p R_{pq}\over 2\pi}
{ik\over R_p^2(\xi_s,\eta_s)R_q(\xi_s,\eta_s)}\
{\partial(x,y)\over\partial(\xi,\eta)}
\hbox{ \ for \ } \Re k\rightarrow\infty,\ \Im k>0               \label{uni1}
\end{equation}
is to be evaluated at the point of
stationarity $\xi_s,\eta_s$ and multiplied by the factor
$2\pi i/k\sqrt{\Delta_{\xi\xi}\Delta_{\eta\eta}-\Delta_{\xi\eta}^2}$
of equation (\ref{exh5}).

Trivially, $R_p$ and $R_q$ are just the two fractions of $R_{pq}$
\begin{equation}
R_p({\xi_s,\eta_s})={z_p\over z_p-z_q}R_{pq}\qquad
R_q({\xi_s,\eta_s})={-z_q\over z_p-z_q}R_{pq}\ .                \label{uni2}
\end{equation}
Distances as they are, they do not depend on the coordinate system.
This is different for the determinant
$\Delta_{\xi\xi}\Delta_{\eta\eta}-\Delta_{\xi\eta}^2$. Let us
calculate it first in Cartesian coordinates. Differentiating
the triangle function (\ref{tri1}) twice and evaluating it
at the stationary point gives
\begin{eqnarray}
\Delta_{xx}={(z_p-z_q)^2((y_p-y_q)^2+(z_p-z_q)^2)\over -z_p z_q R_{pq}^3}\
                                                                  \nonumber\\
\Delta_{yy}={(z_p-z_q)^2((x_p-x_q)^2+(z_p-z_q)^2)\over -z_p z_q R_{pq}^3}\
                                                               \label{uni3}\\
\Delta_{xy}={-(z_p-z_q)^2 (x_p-x_q)(y_p-y_q)\over -z_p z_q R_{pq}^3}\,.
\nonumber
\end{eqnarray}
Consequently
\begin{equation}
\Delta_{xx}\Delta_{yy}-\Delta_{xy}^2={(z_p-z_q)^6\over z_p^2 z_q^2 R_{pq}^4}\ .
                                                                \label{uni4}
\end{equation}
The root of this determinant is transformed multiplying it
by the functional determinant taken also at the point of stationarity
\begin{equation}
\sqrt{\Delta_{\xi\xi}\Delta_{\eta\eta}-\Delta_{\xi\eta}^2}=
\sqrt{\Delta_{xx}\Delta_{yy}-\Delta_{xy}^2}\
{\partial(x,y)\over\partial(\xi,\eta)}\ .                       \label{uni5}
\end{equation}
Additional terms do not occur. They would contain first derivatives
of the triangle function which are zero at the stationary point,
see conditions (\ref{tri14}).

The result is astonishing:
\begin{equation}
{2\pi i\ A(\xi_s,\eta_s)\over k\sqrt{\Delta_{\xi\xi}
\Delta_{\eta\eta}-\Delta_{\xi\eta}^2}}=1+O(k^{-1})          \label{uni6}
\end{equation}
though it should be noted that it is -1 if
the functional determinant is negative, i.e.\ if
the coordinates $\xi$,$\eta$ form a left-handed system.
The final formulae for the representatives derived from (\ref{exh5})
are thus simple and identical
\begin{eqnarray}
a_k({\bf r}_p)={\exp(ikR_{pq})\over R_{pq}}\
{\erfc\sqrt{-ik\Delta_s(\eta_-)} -
 \erfc\sqrt{-ik\Delta_s(\eta_+)}\over 2}+O(k^{-1})\           \label{uni7}\\
b_k({\bf r}_p)={\exp(ikR_{pq})\over R_{pq}}\
{\erfc\sqrt{-ik\Delta_s(\eta_-)}-
 \erfc\sqrt{-ik\Delta_s(\eta_+)}\over 2}+O(k^{-1})\,.         \label{uni8}
\end{eqnarray}

Further simplifaction is possible.
If the upper edge is removed, i.e.\ $\eta_+\rightarrow+\infty$, the
exhausting function is supposed to increase beyond measure
$\Delta_s(\eta_+)\rightarrow+\infty$ and,
according to the asymptotics (\ref{err5}),
the formulae (\ref{uni7}-\ref{uni8}) are reduced to the

\medskip
\noindent{\bf Universal Formula of Diffraction.}
{\sl The diffracted wave is the primary wave times a universal function
describing the change from light to shadow.
\begin{equation}
a\hbox{ or }b_k({\bf r}_p)={\exp(ikR_{pq})\over R_{pq}}\
{\erfc\sqrt{-ik\Delta_s}\over 2}+O(k^{-1})\ .                \label{uni9}
\end{equation}
The geometry of the diffracting edge enters just through the argument
of that function. This argument is calculated
from the exhausting function (\ref{exh2})
evaluated at the position of the edge:
$\Delta_s=\Delta_s(\eta_-)$.
}
\medskip

From here one may return to the apparently more general formulae
above due to the linearity of Maxwell's equations and a superposition
of their solutions. So $\Delta_s$ may be understood as an abbreviation
for $\Delta_s(\eta_-)$ or $\Delta_s(\eta_+)$ as required.
By the way linearity: Factors on the right-hand sides
that do not depend on ${\bf r}_p$ are always free. Your author omitted them.
This is all the more tolerable since the representatives get effective
only if taken together with the supporting vector field ${\bf t}$
in equations (\ref{kir15}) and (\ref{kir16}).
According to (\ref{car}) the tangent vector ${\bf t}$
contains the components $t_x$ and $t_y$. They can be chosen to
adjust strength and polarization of the primary wave according to
experimental givens.

Formula (\ref{uni9}) constitutes the main achievement of this article.

One may check the accuracy of (\ref{uni9})
by insertion into the Helmholtz equations (\ref{hel2}).
One finds that (\ref{uni9}) satisfies
these equations in the order of $k^2$ identically, but
the terms proportional to $k^{3/2}$ cancel only if the exhausting
function $\Delta_s$ satisfies an equation of eikonal type.
\begin{equation}
(\nabla_p\Delta_s)^2=-2{{\bf r}_p-{\bf r}_q\over R_{pq}}\nabla_p\Delta_s\ .
                                                             \label{uni12}
\end{equation}
Indeed, the fulfillment of this equation follows from the definition
of the triangle function (\ref{lig1}) and the condition of utter exhaust
(\ref{exh1}). Hence the largest erroneous term is $O(k)$. But since the order
of the Helmholtz operator is $k^2$, the relative error
is $O(k^{-1})$, as expected.

Checking the fulfillment of the boundary conditions yields at first
sight a worse result, namely a relative error of $O(k^{-1/2})$.
We see this from the asymptotic expansion of the complementary
error function (\ref{err5}).
This asymptotic expression applies because, directly behind the screen,
the probe resides in the shadow where there is, according to (\ref{lig3}),
$\sqrt{\Delta_s}>0$. Thus the argument of the error function
$\sqrt{-ik\Delta_s}$ lies in the right-hand side of the complex plane.
Hence
\begin{equation}
a_k({\bf r}_p)={\exp(ikR_{pq})\over R_{pq}}\
{\exp(ik\Delta_s)\over 2\sqrt{-i\pi k\Delta_s}}+O(k^{-3/2})  \label{uni13}
\end{equation}
instead of an exact zero as required in the boundary conditions (\ref{bou8}).
A similar result holds for the representative $b_k({\bf r}_p)$.
To verify its boundary condition, one must calculate its normal
derivative, i.e.\ its derivative with respect to $z$. While
this derivative generally is $O(k)$, its value on the
screen is $O(k^{1/2})$ such that the relative error is $O(k^{-1/2})$.

At the second view, both results are better than expected.
First, in Kirchhoff's scalar theory of diffraction, the probe must not
approach the screen since Kirchhoff's functions become singular
\cite[p.9]{Kir91}. In the present theory, all expressions stay regular.
This facilitates, as we will in section \ref{sec:imaging},
a simple trick to suppress errors on the boundaries.
Second, formula (\ref{uni9}), even as it is now,
satisfies the boundary conditions not exactly, but accurately.
The reason is the product $k\Delta_s$ in the denominator of (\ref{uni13}).
The exhausting function
$\Delta_s$ measures, in a slightly unusual way, the distance of the probe
from the border of shadow. The distance between that border and the
screen is the largest possible distance available in a diffracting system.
Hence for short waves, both the wave number $k$ and the distance
$\Delta_s$ are big making (\ref{uni13}) negligibly small.

For applications it is worthwile to calculate from (\ref{uni9})
the field strengths according to (\ref{kir15}) and (\ref{kir16}).
For simplicity we let $a_k({\bf r}_p)=0$. Then the electric field
${\bf E}_k({\bf r}_p)$ stems only from the simple curl in (\ref{kir16})
and the magnetic field ${\bf B}_k({\bf r}_p)$ only from the double curl
in (\ref{kir15}).
\begin{eqnarray}
{\bf E}_k({\bf r}_p) {R_{pq}\over\exp(ikR_{pq})}=
{({\bf r}_p-{\bf r}_q)\times{\bf t}\over R_{pq}}\,
{\erfc\sqrt{-ik\Delta_s}\over 2}\,\omega k +                   \nonumber\\
{\nabla_p\Delta_s\times{\bf t}\over\sqrt{\Delta_s}}\,
\exp(ik\Delta_s)\,\sqrt{i\omega^2 k\over 4\pi} + O(k)\quad    \label{uni10}
\end{eqnarray}
\begin{eqnarray}
{\bf B}_k({\bf r}_p){R_{pq}\ \over\exp(ikR_{pq})}=
{({\bf r}_p-{\bf r}_q)(({\bf r}_p-{\bf r}_q){\bf t})-
{\bf t}R_{pq}^2\over R_{pq}^2}\,
{\erfc\sqrt{-ik\Delta_s}\over 2}\,k^2 +                       \nonumber\\
{\nabla_p\Delta_s(({\bf r}_p-{\bf r}_q){\bf t})+
({\bf r}_p-{\bf r}_q+R_{pq}\nabla_p\Delta_s)(\nabla_p\Delta_s{\bf t})
\over R_{pq}\sqrt{\Delta_s}}\,
\exp(ik\Delta_s)\,\sqrt{ik^3\over 4\pi}                       \nonumber\\
+ O(k)\ .\quad                                                \label{uni11}
\end{eqnarray}
The ubiquitous primary wave was drawn to the left-hand sides to direct
attention to the nontrivial terms on the right-hand sides.

We learn from these equations that diffraction of
electromagnetic waves without polarization does not exist.
Yet the contributions to the right-hand sides with
the complementary error function
describe just the vector properties of the primary
wave. The changes of polarization effected by the screen are
described by the contributions with the exponential function.
The contributions with the error function
are proportional to $k^2$ since
$\omega=O(k)$, cf.\ equation (\ref{kom2a}), whereas those with the
exponential function are proportional only to $k^{3/2}$.
Nevertheless one should not be deluded
that polarization by diffraction is always a lower-order
effect. In the shadow the error function decreases dramatically.
It ceases to be $O(1)$ and weakens to be only $O(k^{-1/2})$
as indicated in equation (\ref{uni13}). In the shadow both contributions
to the right-hand sides of (\ref{uni10}) and (\ref{uni11}) reach
comparable sizes. We must expect hitherto unseen effects.

All terms in (\ref{uni10}) and (\ref{uni11}) remain finite
in the entire domain of solution except, of course, immediately at
the edge where Bremmer's effect takes place. The singularities
on the border of shadow caused by $\sqrt{\Delta_s}$
in the denominators are cancelled by the
derivatives $\nabla_p\Delta_s$ in the nominators since
$\Delta_s$ varies quadratically in the vicinity of the border.

For radiation with low frequencies, e.g.\ microwaves,
the electromagnetic fields can be measured directly. However,
the Maxwell equations are real equations for real observables.
Thus we must extract from the fields (\ref{uni10})-(\ref{uni11})
their real parts prior to comparison with experimental data.
In optics with its much higher frequencies,
the classic measurements are all calorimetric ones. One collects
the energy flux impinging on a given surface for times
much longer than inverses of these frequencies.
The observable is here the time average of the Pointing vector ${\bf S}$
\begin{equation}
\bar{\bf S}({\bf r}_p)=\lim_{\tau\rightarrow\infty}{1\over\tau}\int_o^\tau
\Re{\bf E}({\bf r}_p,t)\times\Re{\bf B}({\bf r}_p,t){\hbox{d}t\over\mu}=
{1\over 2\mu}\Re({\bf E}_k({\bf r}_p)\times{\bf B}_k^*({\bf r}_p))\ .
                                                             \label{uni14}
\end{equation}

The relation between time-dependent and time-independent fields is
as defined in equations (\ref{ini1}). The asterisk denotes complex
conjugation; one can affix it either to the magnetic or to the electric
field without affecting the result. The a priori decomposition of the
complex fields (\ref{uni10}-\ref{uni11}) is not needed here.

\section{Diffraction by a straight edge}
\label{sec:edge}
The first application is, of course, diffraction by a straight edge.
For its description, the Cartesian coordinate
system is best. Hence $\xi=x$, $\eta=y$ and
\begin{equation}
-\infty<x<+\infty\qquad -\infty<y<y_-=\hbox{const}\qquad z=0      \label{edg1}
\end{equation}
defines the screen. $y_-$ is the location of the edge.

The exhausting dependence is found from equations (\ref{exh1}) and
(\ref{tri1})
\begin{equation}
\partial_x R_p(x,y)=-\partial_x R_q(x,y)\ .                      \label{edg3}
\end{equation}
Squaring this equation and cancelling identical terms on both sides results
in a quadratic equation for $x$
\begin{equation}
(x_p-x)^2\big((y_q-y)^2+z_q^2)\big)=(x_q-x)^2 \big((y_p-y)^2+z_p^2)\big)\ .
                                                                 \label{edg4}
\end{equation}
$y$ must be considered as given. One solution of this equation is a
mathematical ghost. It appears because the original equation (\ref{edg3})
was squared to get rid of the roots. The other solution, however,
does solve the original. It is the searched-for exhausting dependence
$\xi=\Xi_s(\eta)$ (\ref{exh1}) in Cartesian coordinates:
\begin{equation}
x=X_s(y)={x_p\sqrt{(y_q-y)^2+z_q^2}+x_q\sqrt{(y_p-y)^2+z_p^2}
       \over \sqrt{(y_q-y)^2+z_q^2}+\sqrt{(y_p-y)^2+z_p^2}}\ .  \label{edg5}
\end{equation}
Inserting this dependence in the triangle function (\ref{tri1})
gives the exhausting function of the straight edge
\begin{eqnarray}
\Delta_s(y)=
\sqrt{(x_p-x_q)^2+\big(\sqrt{(y_p-y)^2+z_p^2}+\sqrt{(y_q-y)^2+z_q^2}\,\big)^2}
                                                              \ \nonumber\\
            -\sqrt{(x_p-x_q)^2+(y_p-y_q)^2+(z_p-z_q)^2}\,.      \label{edg6}
\end{eqnarray}
We take this function at the edge $y=y_-$ and need to determine the
sign of $\sqrt{\Delta_s}=\sqrt{\Delta_s(y_-)}$, i.e.\ where there
is shadow or light. Shadow prevails, according to the criterion (\ref{lig4}),
when $y_s<y_-$. With $y_s$ from equation (\ref{tri4}) we find
\begin{equation}
y_p< y_- + (y_- -y_q){z_p\over -z_q}\ .                         \label{edg7}
\end{equation}
Hence equation (\ref{lig3}) reads here
\begin{equation}
\sqrt{\Delta_s}=
\cases{+|\sqrt{\Delta_s}|&if $y_p<y_- - (y_- -y_q)z_p/z_q$ with $z_p>0$,\cr
       -|\sqrt{\Delta_s}|&elsewhere.\cr}                        \label{edg8}
\end{equation}
What remains to be done is to insert this
in the universal formula of diffraction (\ref{uni9}).

\section{Imaging by diffraction}
\label{sec:imaging}
We have now enough experience to perform the simple trick that was
announced in section \ref{sec:universal}. To improve the fulfillment
of boundary conditions, one inserts a mirrored picture of the source
in the same way as Sommerfeld did when he calculated suitable Green
functions, see section \ref{sec:kirchhoff}. If ${\bf r}_q$ is
the position of the source, ${\bf r}_m$ has the same coordinates
$x_q$ and $y_q$, but the opposite value of $z_q$ as explained in (\ref{kir9}).
Let the distances $R_{pq}$ and $R_{pm}$ be defined
according to (\ref{tri3}):
\begin{eqnarray}
R_{pq}=\sqrt{(x_p-x_q)^2+(y_p-y_q)^2+(z_p-z_q)^2}\            \nonumber\\
R_{pm}=\sqrt{(x_p-x_q)^2+(y_p-y_q)^2+(z_p+z_q)^2}\,.          \label{ima1}
\end{eqnarray}
Hence original and mirrored distance become equal,
$\lim R_{pq}=\lim R_{pm}$, if the probe
approaches the screen, $z_p\rightarrow 0$; it does not matter
if the approach takes place in inner or outer space, $z_p>0$ or $z_p<0$.

We must, however, be careful when we introduce the mirrored exhausting
function. In the definition of the original exhausting function taken at
the edge $y=y_-$
\begin{eqnarray}
\Delta_{sq}=\sqrt{(x_p-x_q)^2+
\big(\sqrt{(y_p-y_-)^2+z_p^2}+\sqrt{(y_q-y_-)^2+z_q^2}\,\big)^2}   \nonumber\\
            -\sqrt{(x_p-x_q)^2+(y_p-y_q)^2+(z_p-z_q)^2}            \nonumber\\
\sqrt{\Delta_{sq}}=
\cases{+|\sqrt{\Delta_{sq}}|&if $y_p<y_- -(y_- -y_q)z_p/z_q$ with $z_p>0$,\cr
       -|\sqrt{\Delta_{sq}}|&elsewhere.\cr}                        \label{ima3}
\end{eqnarray}
the assignment of signs of the root belongs to the definition, cf.\
equations (\ref{edg6}) and (\ref{edg8}).
In about three quarters of space, the root of the original exhausting function
is negative, viz.\ for all negative values of $z_p$ and, if $y_p$ is
sufficiently large, for positive values of $z_p$ too. The source lights
the major part of space.

By contrast, the mirrored exhausting function must be defined as
\begin{eqnarray}
\Delta_{sm}=\sqrt{(x_p-x_q)^2+
\big(\sqrt{(y_p-y_-)^2+z_p^2}+\sqrt{(y_q-y_-)^2+z_q^2}\,\big)^2}  \nonumber\\
            -\sqrt{(x_p-x_q)^2+(y_p-y_q)^2+(z_p+z_q)^2}           \nonumber\\
\sqrt{\Delta_{sm}}=
\cases{-|\sqrt{\Delta_{sm}}|&if $y_p<y_- +(y_- -y_q)z_p/z_q$ with $z_p<0$,\cr
       +|\sqrt{\Delta_{sm}}|&elsewhere.\cr}                       \label{ima5}
\end{eqnarray}
Almost everything follows from (\ref{ima3})
replacing $z_q$ with $-z_q$. The only exception is
the assignment of signs of the root. Using the freedom
mentioned in the discussion behind equation (\ref{lig3}),
your author reversed it. In about three quarters of space,
the root of the mirrored exhausting function is positive, viz.\
for all positive values of $z_p$ and, if $y_p$ is sufficiently large,
for negative values of $z_p$ too. Shadow prevails.

Yet immediately over and under the screen,
the signs of original and mirrored functions are the same,
$\lim\Delta_{sq}=\lim\Delta_{sm}$, if the probe
approaches the screen, $z_p\rightarrow 0$; it does not matter
if the approach takes place in inner or outer space, $z_p>0$ or $z_p<0$.

With these functions it is straightforward to construct solutions
that satisfy the boundary conditions (\ref{bou8}) exactly.
Instead of (\ref{uni9}) we obtain
\begin{equation}
a_k({\bf r}_p)=
{\exp(ikR_{pq})\over R_{pq}}\ {\erfc\sqrt{-ik\Delta_{sq}}\over 2} -
{\exp(ikR_{pm})\over R_{pm}}\ {\erfc\sqrt{-ik\Delta_{sm}}\over 2}+O(k^{-1})\
                                                                \label{ima6}
\end{equation}
\begin{equation}
b_k({\bf r}_p)=
{\exp(ikR_{pq})\over R_{pq}}\ {\erfc\sqrt{-ik\Delta_{sq}}\over 2} +
{\exp(ikR_{pm})\over R_{pm}}\ {\erfc\sqrt{-ik\Delta_{sm}}\over 2}+O(k^{-1})\,.
                                                                \label{ima7}
\end{equation}
The symbol $O(k^{-1})$ was retained to remember that the differential
equations (\ref{hel2}) are not exactly fulfilled.

The second terms on the right-hand sides of equations (\ref{ima6})
and (\ref{ima7}) considered isolated appear fantastic. They
describe ghostly radiation from a source at ${\bf r}_m$ that
becomes bright only when it permeates the screen. Yet when they
are considered in cooperation with the first terms, it is recognized
that they describe the unavoidable reflection that is diffracted
in a similar way as the primary wave. The solutions (\ref{ima6})
and (\ref{ima7}) are valid in entire space.

From these solutions it follows that radiation emitted at ${\bf r}_q$
and diffracted by a screen causes a second singularity at ${\bf r}_m$.
In other words, diffraction creates a sharp image. Of course,
the image is as weak as the multiplicative error function indicates,
but it should be observable because the singularity sticks out.

Folks might feel this prediction as daring. They might not
appreciate the immense power of the methods developed here.
The approximate solution (\ref{ima6}-\ref{ima7}) contains,
as a limiting case, the only exact solution of diffraction problems
known so far, namely Sommerfeld's celebrated stringent solution \cite{Som96}.

Sommerfeld's solution deals with the diffraction of a plane wave.
Using the definitions in (\ref{tri3}) and (\ref{ima1}),
$R_{pq}=|{\bf r}_p-{\bf r}_q|$ and $R_{pm}=|{\bf r}_p-{\bf r}_m|$,
the spherical waves in front of the error functions in (\ref{ima6})
and (\ref{ima7}) can be expanded as
\begin{eqnarray}
{\exp(ik|{\bf r}_p-{\bf r}_q|)\over |{\bf r}_p-{\bf r}_q|}
={\exp(ik|{\bf r}_q|)\over |{\bf r}_q|}
  \exp\bigg({ik{-{\bf r}_q\over|{\bf r}_q|}{\bf r}_p}\bigg)
                                       +O(|{\bf r}_q|^{-2})\quad \label{ima8}\\
{\exp(ik|{\bf r}_p-{\bf r}_m|)\over |{\bf r}_p-{\bf r}_m|}
={\exp(ik|{\bf r}_q|)\over |{\bf r}_q|}
  \exp\bigg({ik{-{\bf r}_m\over|{\bf r}_q|}{\bf r}_p}\bigg)
                                          +O(|{\bf r}_q|^{-2})\ .\label{ima9}
\end{eqnarray}
The first factors on the right-hand sides do not depend on the position
of the probe ${\bf r}_p$. They are absorbed in the normalizations of
(\ref{ima6}) and (\ref{ima7}). When the source is moved to infinity,
$|{\bf r}_m|=|{\bf r}_q|\rightarrow\infty$, only
the plane waves remain. Their fronts are defined by the normal vectors
\begin{equation}
{\bf c}_q=\lim_{|{\bf r}_q|\rightarrow\infty}{-{\bf r}_q\over|{\bf r}_q|}
\qquad
{\bf c}_m=\lim_{|{\bf r}_q|\rightarrow\infty}{-{\bf r}_m\over|{\bf r}_q|}
                                                              \label{ima10}
\end{equation}
with directional cosines as components
\begin{equation}
{\bf c}_q={\bf e}_x c_x+{\bf e}_y c_y+{\bf e}_z c_z\qquad
{\bf c}_m={\bf e}_x c_x+{\bf e}_y c_y-{\bf e}_z c_z\ .        \label{ima11}
\end{equation}

Likewise the exhausting function for the plane-wave problem
is defined as the limit $\delta_s(y)=\lim\Delta_s(y)$ for
$|{\bf r}_q|\rightarrow\infty$. The straightforward calculation
transforms (\ref{edg6}) to
\begin{equation}
\delta_s(y)=\sqrt{c_y^2+c_z^2}\sqrt{(y_p-y)^2+z_p^2}-c_y(y_p-y)-c_z z_p\ .
                                                                \label{ima12}
\end{equation}
Taking the value on the edge $y=y_-$ yields the analog of (\ref{ima3})
\begin{eqnarray}
\delta_{sq}=\sqrt{c_y^2+c_z^2}\sqrt{(y_p-y_-)^2+z_p^2}-c_y(y_p-y_-)-c_z z_p
                                                         \nonumber\\
\sqrt{\delta_{sq}}=
\cases{+|\sqrt{\delta_{sq}}|&if $y_p<y_- +z_p c_y/c_z$ with $z_p>0$,\cr
       -|\sqrt{\delta_{sq}}|&elsewhere.\cr}                      \label{ima13}
\end{eqnarray}
and the mirrored exhausting function as analog of (\ref{ima5})
\begin{eqnarray}
\delta_{sm}=\sqrt{c_y^2+c_z^2}\sqrt{(y_p-y_-)^2+z_p^2}-c_y(y_p-y_-)+c_z z_p
                                                         \nonumber\\
\sqrt{\delta_{sm}}=
\cases{-|\sqrt{\delta_{sm}}|&if $y_p<y_- -z_p c_y/c_z$ with $z_p<0$,\cr
       +|\sqrt{\delta_{sm}}|&elsewhere.\cr}                      \label{ima14}
\end{eqnarray}
Thus the analogs of (\ref{ima6}) and (\ref{ima7}) are
\begin{equation}
a_k({\bf r}_p)=
\exp(ik{\bf c}_q{\bf r}_p)\,{\erfc\sqrt{-ik\delta_{sq}}\over 2} -
\exp(ik{\bf c}_m{\bf r}_p)\,{\erfc\sqrt{-ik\delta_{sm}}\over 2}\  \label{ima15}
\end{equation}
\begin{equation}
b_k({\bf r}_p)=
\exp(ik{\bf c}_q{\bf r}_p)\,{\erfc\sqrt{-ik\delta_{sq}}\over 2} +
\exp(ik{\bf c}_m{\bf r}_p)\,{\erfc\sqrt{-ik\delta_{sm}}\over 2}\,.\label{ima16}
\end{equation}
These functions form an {\sl exact description of diffraction\/}. Not only
the boundary conditions are satisfied perfectly, also the Helmholtz
equations (\ref{hel2}) as direct checking shows. One may be surprised
at this, but it becomes comprehensible when it is noticed that in this problem
with the infinite edge and the primary plane wave, the only length is provided
by the position of the probe ${\bf r}_p$. One can introduce new variables
of position instead of ${\bf r}_p$ by scaling the latter with with the
inverse of the wave number $k$. Hence if the solution is known
for one wave number, here for $k\rightarrow\infty$, it is known
for all wave numbers.

Based on this argument, one supplements the proof
of the fact that the two integrals in (\ref{kir11}) and (\ref{kir12})
are equal to $\mp 2\pi/z_pR_{pq}$, respectively. There is no edge at
all. The only length is the distance between source and probe.
One evaluates the integrals for $k\rightarrow\infty$ using stationary
phase according to section \ref{sec:exhaust} and introduces
thereafter scaled variables to show that the result does not depend on $k$.

Sommerfeld's solution is a special case of (\ref{ima15}-\ref{ima16})
for flat incidence, i.e.\ $c_x=0$ thus $\sqrt{c_y^2+c_z^2}=1$.
In this case, $b_k({\bf r}_p)$ is proportional to the magnetic field
${\bf B}_k({\bf r}_p)=-{\bf e}_x k^2 b_k({\bf r}_p)$.
The factor $-k^2$ is swallowed by normalization.
In the problem with the other polarization,
the electric field ${\bf E}_k({\bf r}_p)$ is proportional to
${\bf e}_x a_k({\bf r}_p)$. Therefore Sommerfeld got along without
the representation theorem of section \ref{sec:representatives}.
Moreover, Sommerfeld neither knew the method of stationary phase
for two-dimensional integrals nor did he use the standardized
complementary error function although it does not take much work to show
that Sommerfeld's function is equivalent. Sommerfeld found his function
via an ingenious contour integration, an approach which does not seem
to admit generalization. Finally, it played a role that Sommerfeld
was Felix Klein's pupil. Klein was the most influential advocate of
{\sl uniformization\/} meaning that roots had to be avoided because
of their ambiguities and to be replaced with suitable parametrizations
necessitating transcendental functions. So Sommerfeld introduced,
instead of the roots in (\ref{ima13}) and (\ref{ima14}), artifical
angles, sines and cosines. Probably this {\sl rootophobia\/} is the
reason why Sommerfelds exceedingly important solution - it is
the \lq harmonic oscillator\rq\ of diffraction - was never sufficiently
appreciated and is not presented in most modern textbooks.

Though Sommerfeld's solution holds only for flat incidence $c_x=0$,
it is, due to translational invariance, easily generalized for skew
incidence. This was done long before this article was drafted
\cite[\S~11.6]{Bor97}. Moreover, the solutions
(\ref{ima15}) and (\ref{ima16}) can be conceived as Fourier components
and used to build the most general diffraction at the straight edge.
For example, diffraction of a spherical wave, which is described
by equations (\ref{ima6}) and (\ref{ima7}) only asymptotically,
was calculated by Macdonald \cite{Mac15}. Yet Macdonald's formulae
are so difficult to survey that imaging by diffraction might have
escaped notice. The present approach is simpler and generalizable.

\section{Diffraction by a slit}
\label{sec:slit}
Let us clarify next one of most lugubrious offences of traditional
theory: the difference between Fresnel and Fraunhofer diffraction.
To anticipate the answer, Fraunhofer diffraction is a misnamer. In the
proper sense of {\sl diffraction\/}, Fraunhofer diffraction does not exist.
There is at best {\sl interference\/}. It occurs only
when a primary wave strikes an aperture with at least two opposite edges.
The thus excited secondary waves interfere to generate a pattern
now named Fraunhofer diffraction.
The formulae derived in this article are powerful enough to describe
the gradual transition from Fresnel diffraction to Fraunhofer interference.
All is included in the equations (\ref{uni7}) and (\ref{uni8}).

To understand how the transition proceeds, the simplest example suffices:
(i) Damping negligible; imaginary part $\Im k\rightarrow 0$; the wave number
$k$ is a positive number.
(ii) Diffraction by a straight slit with edges at $y_\pm=\pm y_0$; 2$y_0$
is a positive number meaning the width of the slit.
(iii) Normal incidence of a plane wave; thus $c_x=c_y=0$, $c_z=1$, see
(\ref{ima10}-\ref{ima12}); in equation (\ref{uni8}), the plane wave
replaces the spherical wave; consequently the exhausting function $\Delta_s$
gives way to its descendant $\delta_s$ (\ref{ima12})
\begin{equation}
b_k({\bf r}_p)=\exp(ikz_p)
{\erfc\sqrt{-ik\delta_s(y_-)}-
 \erfc\sqrt{-ik\delta_s(y_+)}\over 2}+O(k^{-1})           \label{slit1}
\end{equation}
with
\begin{eqnarray}
\delta_s(y_\pm)=\sqrt{(y_p\mp y_0)^2+z_p^2}-z_p
                                                            \nonumber\\
\sqrt{\delta_s(y_\pm)}=
\cases{+|\sqrt{\delta_s(y_\pm)}|&if $y_p<\pm y_0$,\cr
       -|\sqrt{\delta_s(y_\pm)}|&elsewhere\ . \cr}          \label{slit2}
\end{eqnarray}
(iv) The representative $a_k({\bf r}_p)$ is disregarded. (v)
Effects of the reflected wave are neglected; $z_p>0$ is implied.

Equations (\ref{slit1}) and (\ref{slit2}) are still powerfull
as the probe can be moved freely within the inner space $z_p>0$.
Especially the equations remain valid when the probe approaches
the screen $z_p\rightarrow 0$. In most classical experiments, however,
the probe is far away from the screen, $z_p\rightarrow\infty$.
Therefore the exhausting function (\ref{slit2}) can be
expanded in terms of inverse powers of $z_p$
\begin{equation}
\delta_s(y_\pm)={y_p^2\mp 2y_p y_0+y_0^2\over 2 z_p}
\bigg(1+O\big( ({y_p\mp y_0\over z_p})^2 \big)\bigg)\ .        \label{slit3}
\end{equation}
It is worthwhile to mention that using the approximated
exhausting function (\ref{exh9}) yields the same result.
The linearization (\ref{exh7}-\ref{exh8}) spoils the approach
to the screen. In the error functions of (\ref{slit1}),
which oscillate quickly, the first two terms on the right-hand side
of (\ref{slit3}) must be retained because only the second term
will yield specific results. But in algebraic expressions,
the leading term will be enough, for example
\begin{equation}
{1\over |\sqrt{\delta_s(\pm y_0)}|}={{\sqrt{2z_p}\over |y_p|}}
\bigg(1+O\big({y_0\over y_p}\big)+
O\big( ({y_p\over z_p})^2 \big)\bigg)\ .                   \label{slit4}
\end{equation}

To pacify simple-minded scientists' dismay at the error functions
in (\ref{slit1}), let us return to elementary functions using the
asymptotic expansion (\ref{err5}). Because of (\ref{slit2}) we must
do it differently for $y_p\rightarrow-\infty$ and $y_p\rightarrow+\infty$.
For the latter case, equation (\ref{err6}) has to be considered before
(\ref{err5}) can be applied
\begin{equation}
\erfc\sqrt{-ik\delta_s(\pm y_0)}\sim 2-
{\exp(ik\delta(\pm y_0))\over\sqrt{-i\pi k}\,|\sqrt{\delta(\pm y_0)}|}
\hbox{ \ for \ }y_p\rightarrow+\infty\ .                  \label{slit5}
\end{equation}
The minus signs from the last line of (\ref{slit2})
and of the argument in (\ref{err6}) cancel.
Opticians interprete this formula as representing the primary wave
by the 2 and a secondary wave radiated from the edge by the exponential
function. Therefore it is a formula for the domain of light.
This seems to be a contradiction to physics since the domain of light
is limited to $|y_p|<y_0$ while we suppose here $y_p\rightarrow+\infty$.
The contradiction is quashed since the two leading 2's cancel in
(\ref{slit1}).

Calculation based on (\ref{slit3}-\ref{slit5}) yields
\begin{equation}
b_k({\bf r}_p)\sim\sqrt{-i}
\exp\bigg(ik\big(z_p+{y_p^2\over 2z_p}\big)\bigg)\sqrt{{2 k y_0^2\over \pi z_p}}
\,{\sin(ky_0 y_p/z_p)\over ky_0 y_p/z_p}\ .                  \label{slit6}
\end{equation}
Notice the sine is brought about by the interference of the
two terms in (\ref{slit1}). The formula is symmetric with respect
to $y_p$. It holds both for positive and negative values.
A simple calculation for $y_p\rightarrow-\infty$
along the lines just sketched confirms this.

For a comparison with experiments, we need the time-averaged
Pointing vector (\ref{uni14}). The probe usually is an absorbing
layer spanned perpendicularly to the energy flux of the primary wave.
The representative in (\ref{slit6})
produces via (\ref{kir15}) and (\ref{kir16}) with ${\bf t}={\bf e}_x$
the electromagnetic fields
\begin{equation}
{\bf B}_k({\bf r}_p)=-{\bf e}_x k^2 b_k({\bf r}_p)\qquad
{\bf E}_k({\bf r}_p)={\bf e}_y\omega k b_k({\bf r}_p)(1+O(z_p^{-1}))
+{\bf e}_z\ ...\ .                                         \label{slit7a}
\end{equation}
The dots behind ${\bf e}_z$ substitute a function we presently
do not need to know. The measured component of the energy flux is thus
\begin{equation}
{\bf e}_z\bar{\bf S}({\bf r}_p)\sim {\omega k^3\over 2\mu}|b_k({\bf r}_p)|^2\ .
                                                               \label{slit7}
\end{equation}
$\bar{\bf S}_0={\bf e}_z\omega k^3/(2\mu)$ is,
because of the normalization chosen in (\ref{slit1}),
the energy flux of the primary wave. Straightforward evaluation of
(\ref{slit7}) with (\ref{slit6}) gives
\begin{equation}
{\bf e}_z\bar{\bf S}({\bf r}_p)\sim |\bar{\bf S}_0|\,2y_0\,{ky_0\over \pi z_p}
\bigg({\sin(ky_0 y_p/z_p)\over ky_0 y_p/z_p}\bigg)^2\ .        \label{slit8}
\end{equation}
This is the most fabulous formula of traditional diffraction theory -
Kirchhoff's formula of Fraunhofer {\sl diffraction\/}.
However, Kirchhoff found only the factor with the squared sine.
Historically, various normalizing factors were adjusted a posteriori, but
the fabulous formula was never perceived as an equation among vectors.

Here, by contrast, everything arises from first principles.
The last two factors in (\ref{slit8}) when integrated over $y_p$ from
$-\infty$ to $+\infty$ yield 1. The first two factors describe
the primary energy flux squeezed through the slit of width $2y_0$.
Equation (\ref{slit8}) expresses energy conservation as warranted
by Maxwell's equations in a medium without damping.

The fabulous formula is worse than admitted in textbooks.
From the estimate in (\ref{slit4}) we see it holds only for
\begin{equation}
y_0<<|y_p|<<z_p\ .                                      \label{slit9}
\end{equation}
$y_0<<z_p$ means that the probe is far away from the screen.
This is all right because $z_p\rightarrow\infty$ was announced at the
start of the calculation. But the extendend condition (\ref{slit9})
entails that Kirchhoff's formula provides
neither a valid description of the central peak around
$|y_p|\le y_0$ nor of large-angle diffraction $|y_p|>z_p$.
Only some intermediate wiggles are correctly seized.
That the condition $y_0<<|y_p|$ matters, is corroborated through
the use of the asymptotic expansion (\ref{slit5}).
It is utterly wrong at values of the exhausting function close
to zero. The triangle function and thus the exhausting function of
optics is, according to (\ref{lig2}), zero on the border of shadow
defined here by $y_0=|y_p|$.

The positive outcome of the preceding discussion
is better understanding of the working of formula
(\ref{slit1}) and its prototypes (\ref{uni7}) and (\ref{uni8}).
Let us position the probe first close to an edge. To describe
this measurement, one of the terms in (\ref{slit1}) is enough.
We must, however, utilize the error function full-fledged.
When the probe is removed from the edge, but kept close
to the screen, we are still in the realm of pure diffraction.
Still one term of (\ref{slit1}) suffices, but we may use now its
asymptotic expansion (\ref{err5}). Only in that part of space
where distances to the edges are about equal, interferences
show up. Both terms in (\ref{slit1}) must be kept.
When the distances become large, we may use asymptotic expansions,
but not at the borders of shadow.
With increasing distances from the edges, the diffracted beams spread.
In particular this is true close to the borders. The central beam
is enclosed between these borders. Therefore it is always questionable
to decribe diffraction and interference in the vicinity of
the central beam using the asymptotic expansion (\ref{err5}).

\section{Diffraction by a circular aperture}
\label{sec:circular}
After it is confirmed that the new theory encloses correct parts
of the traditional theory as special cases, we are free to step into the
entirely unknown.

The round stop is the most often installed component
in optical instruments. We want to calculate the diffraction it causes,
more precisely its Fresnel diffraction, since some information on Fraunhofer
interference is already known - Airy's formula \cite[\S~8.5.2]{Bor97}.
The edge of a circular aperture is not straight. Nevertheless the
universal formula of diffraction (\ref{uni9}) applies again.
All we have to do is to find the suited exhausting function.

The suitable coordinate system is the cylindrical one introduced in section
\ref{sec:triangle}, see (\ref{tri5}).
\begin{equation}
-\pi <\varphi\le +\pi\qquad 0\le \rho<\rho_0=\hbox{const}\qquad z=0 \label{cir1}
\end{equation}
defines a circular aperture in the screen.
The exhausting dependence is found from equations (\ref{exh1}) and
(\ref{tri1})
\begin{equation}
\partial_\varphi R_p(\rho,\varphi)=-\partial_\varphi R_q(\rho,\varphi)
                                                                \label{cir2}
\end{equation}
with
\begin{eqnarray}
R_p(\rho,\varphi)=\sqrt{\rho_p^2+\rho^2-2\rho_p\rho
                        \cos(\varphi_p-\varphi)+z_p^2}\       \nonumber\\
R_q(\rho,\varphi)=\sqrt{\rho_q^2+\rho^2-2\rho_q\rho
                        \cos(\varphi_q-\varphi)+z_q^2}\,.     \label{cir2a}
\end{eqnarray}
For normal incidence, $\rho_q=0$, one can observe the solutions
using the rubber-ribbon experiment explained in section \ref{sec:exhaust}.
Actually, there are two:
\begin{equation}
\varphi=\varphi_p\hbox{ \ and \ }\varphi=\varphi_p+\pi\ .     \label{cir2b}
\end{equation}
These exhausting dependencies $\varphi=\Phi_s(\rho)$,
analogues of $\xi=\Xi_s(\eta)$ in (\ref{exh1}), surprise as they
do not depend on $\rho$. Inserting them into the triangle function
(\ref{tri1}) gives the exhausting function of the circular stop.
We need it at the edge $\rho=\rho_0$:
\begin{equation}
\Delta_s=\sqrt{(\rho_p-\rho_0)^2+z_p^2}+\sqrt{\rho_0^2+z_q^2}
-\sqrt{\rho_p^2+(z_p-z_q)^2}\ .                               \label{cir2c}
\end{equation}
The definition
has to be completed with the disposal of the root
\begin{equation}
\sqrt{\Delta_s}=
\cases{+|\sqrt{\Delta_s}|&if $\rho_p>\rho_0(z_q-z_p)/z_q$ with $z_p>0$,\cr
       -|\sqrt{\Delta_s}|&elsewhere,\cr}                     \label{cir2d}
\end{equation}
according to equations (\ref{lig3}) and (\ref{tri7}).
The exhausting function of the second solution in (\ref{cir2b})
is derived from this replacing $\rho_0$ with $-\rho_0$.
Your author names the first solution with the positive $\rho_0$
{\sl near\/} because its point on the edge is closer to the probe
than in the opposite case. The second solution with the negative $\rho_0$
is thus called {\sl far\/}.

For mathematical reasons, only the near solution may dominate.
The definition (\ref{cir1}) of cylindrical coordinates restricts
the range of the angle $\varphi$. Yet unrestricted variation
of the variable $\xi$, analog of $\varphi$, is a premise
of the principle of utter exhaust as declared in section
$\ref{sec:exhaust}$. The violation of this premise is not disastrous
as long as the exhausting function $\Delta(\xi,\eta)$ takes a sharp extremum
on the edge. In this case, the finiteness of integration causes
negligible corrections. But for the round stop $\Delta(\xi,\eta)$
becomes equal for all points on the edge when source and probe
both approach the optical axis;
there is no extremum of $\Delta(\xi,\eta)$ at all. Consequently the
universal formula of diffraction (\ref{uni9}) applied to circular
apertures needs $\rho_p>z_p>0$ to secure its applicability.

Of course, the primary wave is also diffracted at the far side of the
circular stop. There will be similar interferences as those caused by
the slit, see section \ref{sec:slit}, but the contributions from the
far side are here, in the range of validity just declared, small.
Fraunhofer interferences cannot be described well.
Nevertheless, within the restricted range, there seems to be no rival
of the formulas (\ref{cir2c}) and (\ref{cir2d}).
We can calculate diffraction immediately behind a circular aperture
and predict what a curved edge does to polarization.

For general parameters, the exhausting equation (\ref{cir2}) cannot
be solved in terms of roots. When the transcendental functions sine
and cosine are rationalized introducing, for instance, the variables
$c$ and $s$
\begin{equation}
c=\cos\big(\varphi-{\varphi_p+\varphi_q\over 2}\big)\qquad
s=\sin\big(\varphi-{\varphi_p+\varphi_q\over 2}\big)            \label{cir4}
\end{equation}
where $c^2+s^2=1$, and when the roots are removed by squaring,
an algebraic equation of the sixth degree is obtained.

Nevertheless we can find more physically relevant solutions,
even for oblique incidence, performing the rubber-ribbon experiment as above.
Namely the {\sl one-sided flat situation\/}
where source and probe sit on the same
side of the optical axis, and the {\sl two-sided flat situation\/}.
In both situations, there are near and far solutions:
\begin{eqnarray}
\hbox{one-sided flat: }\varphi_p=\varphi_q\phantom{+\pi\ }\hbox{ \ near: }\varphi=\varphi_p
\hbox{ \ far: }\varphi=\varphi_p+\pi\                     \label{cir4a}\\
\hbox{two-sided flat: }\varphi_p=\varphi_q+\pi \hbox{ \ near: }\varphi=\varphi_p
\hbox{ \ far: }\varphi=\varphi_p+\pi\,.                   \label{cir4b}
\end{eqnarray}
But this is not enough when vector fields as in (\ref{uni10})
and (\ref{uni11}) are to be calculated. For them
we must differentiate the exhausting function with respect to
$z_p$, $\rho_p$ and $\varphi_p$. The last differentiation is not
feasible when the function is only known for a fixed $\varphi_p$.

Mending the shortcoming is easy.
The sine $S$ and the cosine $C$ were defined in (\ref{tri8}).
In the context of (\ref{cir2}), they must be considered as given and fixed.
The sine $s$ and the cosine $c$ were defined in (\ref{cir4}). They contain
the angle which is sought as the exhausting dependence
$\varphi=\Phi_s(\rho)$. With the abbreviations
\begin{equation}
P(\rho)=\sqrt{(\rho_p-\rho)^2+z_p^2}/\rho_p\qquad
Q(\rho)=\sqrt{(\rho_q-\rho)^2+z_q^2}/\rho_q                     \label{cir5}
\end{equation}
the exhausting equation (\ref{cir2})
can be written in four equivalent variants
\begin{equation}
 {Cs-Sc\over \sqrt{P(\pm\rho)^2-2\rho(Cc+Ss\mp 1)/\rho_p}}=
-{Cs+Sc\over \sqrt{Q(\pm\rho)^2-2\rho(Cc-Ss\mp 1)/\rho_q}}\ .   \label{cir3}
\end{equation}
Differently to all other equations in this article,
signs can be chosen independently. Just under the roots,
the signs must be altered coherently.

These variants permit to find solutions in the neighborhood
of (\ref{cir4a}) and (\ref{cir4b}) such that they are exact in linear terms.
From (\ref{tri8}) we have

\begin{tabular}{rll}
one-sided: &$S=O(\varphi_p-\varphi_q)$ &$C= 1+O((\varphi_p-\varphi_q)^2)$\\
two-sided: &$S=1+O((\varphi_p-\varphi_q-\pi)^2)$ &$C=O(\varphi_p-\varphi_q-\pi)$\ .\\
\end{tabular}

\noindent Inserting the estimates according to (\ref{cir4a}-\ref{cir4b})

\begin{tabular}{rl}
one-sided near: &$\varphi=\varphi_p+O(\varphi_p-\varphi_q)$\\
one-sided far:  &$\varphi=\varphi_p+\pi+O(\varphi_p-\varphi_q)$\\
two-sided near: &$\varphi=\varphi_p+O(\varphi_p-\varphi_q-\pi)$\\
two-sided far:  &$\varphi=\varphi_p+\pi+O(\varphi_p-\varphi_q-\pi)$\\
\end{tabular}

\noindent in the definitions (\ref{cir4}) yields

\begin{tabular}{rll}
one-sided near: &$s=O(\varphi_p-\varphi_q)$ &$c= 1+O((\varphi_p-\varphi_q)^2)$\\
one-sided far:  &$s=O(\varphi_p-\varphi_q)$ &$c=-1+O((\varphi_p-\varphi_q)^2)$\\
two-sided near: &$s= 1+O((\varphi_p-\varphi_q-\pi)^2)$ &$c=O(\varphi_p-\varphi_q-\pi)$\\
two-sided far:  &$s=-1+O((\varphi_p-\varphi_q-\pi)^2)$ &$c=O(\varphi_p-\varphi_q-\pi)\ .$\\
\end{tabular}

\noindent Hence the nominators in (\ref{cir3}) are always small
of first order. In the denominators,
$P(\pm\rho)$ and $Q(\pm\rho)$ are positive constants. But the expressions
$Cc+Ss\mp 1$ and $Cc-Ss\mp 1$ are always small of second order if
signs are suitably selected. Thus these expressions can be neglected.
Without extertion one finds now the exhausting dependencies
for the one-sided flat situation
\begin{equation}
\hbox{near: \ }    s=-S\,{P( \rho)-Q( \rho)\over P( \rho)+Q( \rho)}\qquad
\hbox{ \ far: \ }  s= S\,{P(-\rho)-Q(-\rho)\over P(-\rho)+Q(-\rho)}\label{cir6}
\end{equation}
and for the two-sided flat situation
\begin{equation}
\hbox{near: \ }   c=-C\,{P( \rho)+Q(-\rho)\over P( \rho)-Q(-\rho)}\qquad
\hbox{ \ far: \ } c= C\,{P(-\rho)+Q( \rho)\over P(-\rho)-Q( \rho)}\ .\label{cir7}
\end{equation}
The exhausting function of the one-sided flat situation, based on the near
solution and taken at the edge $\rho=\rho_0$, is, when abbreviations are
restored,
\begin{eqnarray}
\Delta_s=\sqrt{(\rho_p-\rho_0)^2+z_p^2}+\sqrt{(\rho_q-\rho_0)^2+z_q^2}
\qquad\qquad                                                         \nonumber\\
                -\sqrt{(\rho_p-\rho_q)^2+(z_p-z_q)^2}\qquad\qquad    \nonumber\\
 +2\,\bigg(\big({\sqrt{(\rho_p-\rho_0)^2+z_p^2}\over\rho_p\rho_0}+
       {\sqrt{(\rho_q-\rho_0)^2+z_q^2}\over\rho_q\rho_0}\big)^{-1}\qquad\nonumber\\
 -\big({\sqrt{(\rho_p-\rho_q)^2+(z_p-z_p)^2}\over \rho_p\rho_q}\big)^{-1}\bigg)
   \sin^2{\varphi_p-\varphi_q\over 2}                              \label{cir8}
\end{eqnarray}
$\sqrt{\Delta_s}=+|\sqrt{\Delta_s}|$ \ if
\begin{equation}
\rho_p>{\rho_0(z_q-z_p)+\rho_q z_p\over z_q}
(1+2{\rho_q z_p\over\rho_0(z_p-z_q)}\sin^2{\varphi_p-\varphi_q\over 2})
\hbox{ with }z_p>0,                                           \label{cir8a}
\end{equation}
$\sqrt{\Delta_s}=-|\sqrt{\Delta_s}|$ \ elsewhere.

\medskip\noindent
The equations (\ref{tri7}-\ref{tri8}) were used to determine the border
of shadow.

The respective expressions the two-sided flat situation are
\begin{eqnarray}
\Delta_s=\sqrt{(\rho_p-\rho_0)^2+z_p^2}+\sqrt{(\rho_q+\rho_0)^2+z_q^2}
\qquad\qquad                                                        \nonumber\\
                -\sqrt{(\rho_p+\rho_q)^2+(z_p-z_q)^2}\qquad\qquad   \nonumber\\
 +2\,\bigg(\big({\sqrt{(\rho_p-\rho_0)^2+z_p^2}\over\rho_p\rho_0}-
       {\sqrt{(\rho_q+\rho_0)^2+z_q^2}\over\rho_q\rho_0}\big)^{-1}\quad\nonumber\\
 +\big({\sqrt{(\rho_p+\rho_q)^2+(z_p-z_p)^2}\over \rho_p\rho_q}\big)^{-1}\bigg)
   \cos^2{\varphi_p-\varphi_q\over 2}                              \label{cir9}
\end{eqnarray}
$\sqrt{\Delta_s}=+|\sqrt{\Delta_s}|$ \ if
\begin{equation}
\rho_p>{\rho_0(z_q-z_p)-\rho_q z_p\over z_q}
(1-2{\rho_q z_p\over\rho_0(z_p-z_q)}\cos^2{\varphi_p-\varphi_q\over 2})
\hbox{ with }z_p>0,                                           \label{cir9a}
\end{equation}
$\sqrt{\Delta_s}=-|\sqrt{\Delta_s}|$ \ elsewhere.

\medskip\noindent
All exhausting functions based on the far solutions can be obtained
from these formulas inverting the sign of $\rho_0$.
The errors are $O((\varphi_p-\varphi_q)^4)$ or
$O((\varphi_p-\varphi_q-\pi)^4)$, respectively.

Generally, one may solve equation (\ref{cir3}) numerically. The effort
for this is orders of magnitudes less than for a direct numerical
solution of Maxwell's equations.

\section{Outlook}
\label{sec:outlook}
As we all know, everthing was foretold by Aristotle
or Jesus Christ. So it is only with timidly throbbing heart
that your authors dares to ask the deciding question:
What might be novel in this article?

All monographies and textbooks pretending to teach the solution
of Max\-well's equations should be committed to paper recycling.
The same applies to those treatises wherein optics is allegedly
derived from electrodynamics. The general representation theorem
in section \ref{sec:representatives} beats them all.
That the boundary conditions on metallic surfaces are as simple
for the representatives as deduced in section \ref{sec:boundary}, this
was never anticipated. These discoveries are synthesized with
Sommerfeld's criticisms and proposals in the sections \ref{sec:initial}
and \ref{sec:kirchhoff}. The thus derived solution of
electromagnetic diffraction polarization inclusive,
described in section \ref{sec:kirchhoff}, is a new finding.
Compared to this, the discussion of the triangle function
in section \ref{sec:triangle} appears at first sight of minor importance.
But the root of the triangle function and a consistent assignment
of its signs are of prime importance for applications.
The big thing, however, is the principle of utter exhaust
declared in section \ref{sec:exhaust}. For the first time,
stationary phase is applied where it is really needed. Utter exhaust
seems to have no predecessors at all. It is so powerful,
it would be explained in dozens of textbooks if it was known.
Utter exhaust shows its power in the universal formula of diffraction,
see section \ref{sec:universal}. Its application to diffraction
by an edge in section \ref{sec:edge} yields a comprehensive formula
which encloses, as a limiting case,
Sommerfeld's stringent solution, see section \ref{sec:imaging}.
The great wonder is the perfect agreement though utter exhaust
is just an asymptotic evaluation for short wavelengths. For
diffraction by a slit, as discussed in section \ref{sec:slit},
utter exhaust copes with Fresnel diffraction
and Fraunhofer interference in a unified way. The gradual transition
between these fundamentally different patterns can be calculated.
The derivation of these formulae and their notation is new.
Everything is now easier to survey than in previous theories
The formulas in section \ref{sec:circular} describing diffraction
by a round stop, derived via utter exhaust, are completely new.

Hundreds of physicists will find rewarding foundations for
future business. The physical discussion of the formulas derived here
has only begun. Important systems, e.g.\
diffraction by a hook or a cusp, were not yet analyzed
though they are now accessible. Moreover, the circular aperture
is an especially rich system which deserves further studies.
The foundation of its theory is the integral (\ref{kir14}).
Here it was evaluated for short wavelengths only, but more
asymptotics must be considered. Evaluation for large distances
from the screen will yield Fourier integrals and thus comfort all those
who understand imaging as a sequence of Fourier and inverse Fourier
transforms. The asymptotics for small angles are scarcely understood
though they are most important for applications in optical systems.
Quantitative comparisons with experiments are altogether rare.
The best measurements are probably those made with microwaves \cite{Kin59}.

At last we have a reasonable theory of electrodynamic diffraction,
but also part of acoustics will change. The methods developed
from section \ref{sec:boundary} on apply to the velocity field of sound
if the representatives $a_k({\bf r}_p)$ and $b_k({\bf r}_p)$ are construed
as scalar potentials. The boundary conditions derived in section
\ref{sec:boundary} are just the most often used ones in acoustics,
namely on the soft or the hard wall, i.e.\ acoustic impedance
zero or infinite. Almost everything remains the same, in particular
the universal formula of diffraction (\ref{uni9}).
Only the dispersion relation of sound is different from (\ref{kom1})
since compressible fluids refuse the telegraph equation \cite{Bro86}.

It is time to reconcile with Gustav's fans. Your author knows
the pitfalls of pioneering. He reveres Gustav Kirchhoff affectionately
and repents bitterly not to have read his books first.
Almost all epigone scientists copied Kirchhoff's ideas,
and it is sometimes felt that the copies were made without
understanding. In other words, modern scientists do not attain
Kirchhoff's intellectual level. It is, however,
remarkable that just Max Born copied Kirchhoff's. Born is reputed
as the founder of quantum mechanics in so far as he invented the
probabilistic interpretation. A careful comparison of Born's
monography \cite{Bor65} with Kirchhoff's textbook on optics \cite{Kir91}
revealed that Born copied the probabilistic interpretation, too.
Kirchhoff, an adherent of German idealistic philosophy
according to which material objects descend from ideas, made a theory
for {\it Verr\"uckungen\/}, i.e.\ for nonobservable items. The reader
might remark the contrast to Maxwell who based his theory on measurable
quantities, namely on the electromagnetic fields. Nevertheless Kirchhoff,
being also a physicist, desired to compare his ideas to measurements.
So he stipulated, without any reason, that the squares
of those {\it Verr\"uckungen\/} were {\it intensities\/},
i.e.\ nonnormalized probabilities:
{\lq\lq}Von dieser h\"angt die Intensit\"at J des Lichtes ab.
Als Maass derselben f\"uhren wir den Mittelwerth des Quadrates
der Verr\"uckung eines Aetherteilches ein{\rq\rq} \cite[p.9]{Kir91}.
Kirchhoff is the true inventor of the probabilistic interpretation,
but the invention was not a sign of mental strength, but lack
of ability to solve the correct differential equations correctly.
As shown in this article, in particular in section \ref{sec:slit},
electrodynamics do not need a {\sl probabilistic interpretation\/}.

The Dirac equation, actually a system of coupled partial differential
equations, does the same for electronic motion as Maxwell's equations
do for photonic propagation. A probabilistic interpretation
is not necessary because the Dirac equation describes observables,
namely the external electrical charge and current densities
which were discussed in section \ref{sec:representatives}.
The Dirac equation even defies that interpretation. Eighty years
after Dirac's discovery, no compatible probability formalism
for the Dirac equation is known.

Your author meanwhile derived representation theorems for the
Dirac equation \cite{Bro10}. They simplify its solution in the same way
as the representation theorem of section \ref{sec:representatives}
simplifies the solution of Maxwell's equations.
Schr\"odinger's $\psi$ turns out to be a representative similar
to the functions $a$ and $b$ used here. The Schr\"odinger equation
appears as an auxiliary similar the telegraph equation.
Interpreting Schr\"odinger's mathematical tools physically
appears as {\it Verr\"uckung\/} again.

Moreover it is already known that several effects seeming to prove
probabilistic behavior in quantum electrodynamics can be derived
instead from the spatial distributions of charge and current
as determined by the Dirac equation \cite{Bar83}.
Nevertheless, a certain group of experiments exhibits random behavior
of photons and electrons, viz.\ absorption by small absorbers \cite{Ton89}.
Yet the respective dynamical problems were never solved -
as pipe turbulence was not solved before 1988 \cite{Bob88}.
Random behavior in quantum electrodynamics might stem
from deterministic chaos similarly as in other areas of nature.
It might happen that Heisenberg's mysticism regarding quantum
{\lq\lq}particles{\rq\rq} and {\lq\lq}uncertainty{\rq\rq}
will be gone soon.

\end{document}